\documentclass[fleqn,usenatbib]{mnras}

\usepackage{newtxtext, newtxmath}

\usepackage[T1]{fontenc}
\usepackage{ae, aecompl}

\usepackage{amsmath}            % Advanced maths commands
\usepackage{amssymb}            % Extra maths symbols
\usepackage{graphicx}           % Including figure files
\usepackage[utf8]{inputenc}     % UTF-8 encoding for input (needed for bib)
\usepackage{textgreek}          % Greek letters in text mode
\usepackage{siunitx}            % Unit typesetting
\newcommand{\crash}{\textsc{crash}}
\newcommand{\diff}[1][\empty]{\ifx\empty#1
                \ensuremath{\mathop{}\!\mathrm{d}}
        \else
                \ensuremath{\mathop{}\!\mathrm{d}^#1}
        \fi}
\newcommand{\tder}[3][\empty]{\ifx\empty#1
                \ensuremath{\frac{\diff #2}{\diff #3}}
        \else
                \ensuremath{\frac{\diff[#1] #2}{\diff #3^{#1}}}
        \fi}

\newcommand{\HI}{{\text{H}\,\textsc{i}}}
\newcommand{\HII}{{\text{H}\,\textsc{ii}}}
\newcommand{\HeI}{{\text{He}\,\textsc{i}}}
\newcommand{\HeII}{{\text{He}\,\textsc{ii}}}
\newcommand{\HeIII}{{\text{He}\,\textsc{iii}}}

\sffamily % load T1lmss.fd
\DeclareFontShape{T1}{aer}{bx}{sc} { <-> ssub * cmr/bx/sc }{}
\title[dustyCRASH]{Radiative transfer of ionizing radiation through gas and
  dust: stellar source case}

\author[M. Glatzle et al.]{
Martin Glatzle,$^{1,2}$\thanks{E-mail: mglatzle@mpa-garching.mpg.de}
Benedetta Ciardi$^{1}$ and Luca Graziani$^{3,4}$
\\
$^{1}$Max Planck Institut f\"ur Astrophysik, Karl-Schwarzschild-Str. 1,
85748 Garching, Germany\\
$^{2}$Physik-Department, Technische Universit\"at M\"unchen,
James-Franck-Str. 1,
85748 Garching, Germany\\
$^{3}$INAF Osservatorio Astronomico di Roma, Via Frascati 33, 00040, Monte
Porzio Catone (RM), Italy \\
$^{4}$ Scuola Normale Superiore, piazza dei Cavalieri 7, I-56126 Pisa, Italy}

\date{Accepted XXX. Received YYY; in original form ZZZ}

\pubyear{2017}

\begin{document}
\label{firstpage}
\pagerange{\pageref{firstpage}--\pageref{lastpage}}
\maketitle

\begin{abstract}
  We present a new dust extension to the Monte Carlo radiative transfer code
  \crash , which enables it to simulate the propagation of ionizing radiation
  through mixtures of gas and dust. The new code is applied to study the
  impact of dust absorption on idealized galactic \HII\ regions and on small
  scale reionization. We find that \HII\ regions are reduced in size by the
  presence of dust, while their inner temperature and ionization structure
  remain largely unaffected. In the small scale reionization simulation, dust
  hardens ionization fronts and delays the overlap of ionized bubbles. This
  effect is found to depend only weakly on the assumed abundance of dust in
  underdense regions.
\end{abstract}

\begin{keywords}
  radiative transfer -- ISM: dust, extinction -- cosmology: dark ages,
  reionization, first stars
\end{keywords}

%%%%%%%%%%%%%%%%%%%%%%%%%%%%%%%%%%%%%%%%%%%%%%%%%%

%%%%%%%%%%%%%%%%% BODY OF PAPER %%%%%%%%%%%%%%%%%%

\section{Introduction}

The existence of cosmological dust in the diffuse Galactic interstellar
medium (ISM) was first inferred from the extinction of starlight
\citep{Trumpler1930}. Since that time, much observational evidence of dust
has been found, for example in star-forming regions
\citep[e.g.][]{Churchwell2009} or our Solar System \citep[e.g.][]{Srama2011},
and it has become increasingly apparent that, at least in collapsed
structures, dust is ubiquitous in the Local Universe.

One of the best-studied properties of dust is its wavelength dependent
extinction \citep[e.g.][]{Fitzpatrick1999}. Although the exact dependence
varies from environment to environment, a very general finding is that dust
preferentially absorbs visible and ultra-violet (UV) light
\citep[e.g.][]{Draine2011}. Though observationally less well constrained,
models predict a large dust cross section also in the ionizing UV
(\SIrange{13.6}{200}{eV}; see \S~\ref{sec:silic-graph-pah}).

The former point has important consequences for the observation of recent
star formation, traced by massive and short-lived stars, which emit most of
their energy in the UV (e.g. \citealt{Kennicutt2012}). These stars are
observed to form in groups (OB associations) in dense, dusty molecular clouds
\citep[e.g.][]{Garay1999, deZeeuw1999}, so that a large fraction of their
radiation is often absorbed by dust, which re-emits in the infra-red (IR),
impeding direct observation of the star formation process and strongly
changing galaxy colours \citep[e.g.][]{Mancini2016, Lewis2017}.

The latter point has interesting implications for the formation of \HII\
regions by ionizing UV radiation, one of the feedback channels of massive
stars on the ISM. Despite a long series of studies investigating their
spectral line diagnostics; structure, evolution and physical status of the
ionized gas in these regions still remain subjects of intense theoretical and
observational activity (see e.g. \citealt{Odell2001, Whitney2004,
  Purcell2009, Winston2011, ODell2017}). Galaxies of the Local Group offer a
wide range of \HII\ regions accessible at a level of detail comparable to
Galactic observations \citep{Kennicutt1984, Massey1998}, but located in
differing ISM environments, e.g. with quiescent star formation as observed in
M31 \citep{Azimlu2011} or M33 \citep{Relano2016}. Beyond the Local Group a
large statistical sample of star forming galaxies can also be studied at the
price of resolving less details, but with the advantage of investigating a
wider variety of active star forming environments \citep{Gilbert2007}.

The connection between the size of \HII\ regions and their electron number
density (the so called size-density relation) in observational samples has
been investigated as tracer of the physical status and structure of their ISM
(see \citealt{Hunt2009, Draine_2011} and references therein). Dust certainly
plays a role in shaping this relation and has been recognized as crucial to
its deviation from pure Str\"omgren sphere \citep{Stroemgren1939}
expectations (e.g. \citealt{Inoue2001, Arthur2004}).

Understanding dusty \HII\ regions is therefore important in the context of star
formation; it is, however, also closely linked to understanding the
reionization of the Universe. The importance of the role galaxies played in
this process depends on what fraction $f_\mathrm{UV}$ of the ionizing UV
radiation they produce escapes to the intergalactic medium (IGM)
\citep{Madau1999, Meiksin2009, firstGal2013}. Dust presumably influenced
$f_\mathrm{UV}$, since several galaxies observed in the epoch of reionization
(EOR) appear to be dusty \citep[e.g.][]{Watson2015, Willott2015,
  Laporte2017}, indicating that grain populations were present already during
the EOR.

3D radiative transfer (RT) simulations are the most powerful tool available
to theorists to investigate this topic. Very capable codes have been written
and, in some cases, published for dust RT \citep[e.g.][]{Steinacker2013,
  Gordon2017}, gas RT \citep[which recently is often coupled to
hydrodynamics, e.g.][]{Bisbas_2015} and gas RT on cosmological scales
\citep[e.g.][]{Iliev2006, Iliev2009}\footnote{Note that, since the literature
  for each of these areas is vast, we only cite a few reviews or comparison
  papers.}. However, while the spectral synthesis code \textsc{Cloudy}
\citep{Ferland2017}, or other photon dominated region codes \citep[for a
comparison of several of them see][]{Roellig2007} have included dust effects
for many years, dust and gas are generally not treated together in 3D RT
codes. We are aware of the following exceptions: \textsc{mocassin}
\citep{Ercolano2005}, the code used in \cite{Wood2010}, \textsc{art$^2$}
\citep{Yajima2012}, \textsc{sedna} \citep[\S~3.9]{Bisbas_2015} and
\textsc{torus-3dpdr} \citep{Bisbas2015}, all of which are gas RT codes that
feature dust physics and chemistry models of varying complexity. Recently,
gas and dust RT have also been coupled in the modelling of protoplanetary
discs. Taking advantage of the symmetry of the problem, these codes are,
however, generally 2D \citep[e.g][]{Woitke2009, Bruderer2014}.

As explained above, dust is often present where ionizing photons are
produced, and it will compete with the gas for their absorption (see also
\S~\ref{sec:silic-graph-pah}). The emission of photoelectrons from dust can
significantly contribute to gas heating \citep{Draine2011} and introduces a
direct coupling of gas and dust via a shared free electron pool. We therefore
believe that it is important to be able to treat RT through gas and dust
simultaneously.

Here we perform a first step towards this goal by applying a newly extended
RT code to model the impact of the ionizing radiation emitted by a single
young O--type star in the Local Universe on a dust-polluted
medium\footnote{Analytic solutions for this problem have been published in
  the literature \citep[e.g.][]{Petrosian1972} and we compare our code
  against them in Appendix~\ref{sec:comparison-petrosian}.}. In this context
we also evaluate the contribution of PAHs to dust absorption at ionizing
photon energies. We furthermore investigate the effect of dust in the early
Universe on small scale reionization by artificially polluting a small cosmic
web. This is done in preparation of future applications we envision for our
code, one of which is studying the impact of dust production on the evolution
and detectability of high redshift objects and their surroundings. For the
time being, we consider only the absorption of ionizing photons by dust.

This paper is organized as follows: in \S~\ref{sec:crash} we describe \crash
, the 3D RT code used to simulate ionization of hydrogen and helium gas,
which we extended by a dust module. In \S~\ref{sec:dust-models} we describe
the dust model we use and provide details on the
implementation. \S~\ref{sec:results} contains our results and in
\S~\ref{sec:conclusions} we present our conclusions.

\section{\crash}
\label{sec:crash}
\crash\ (Cosmological RAdiative transfer Scheme for Hydrodynamics) is a 3D
code developed specifically to study time dependent radiative transfer (RT)
problems in cosmology. The Monte Carlo (MC) algorithm it adopts enables
accurate modelling of radiation-matter interactions (generally for photon
energies $h\nu\geq \SI{13.6}{eV}$) and allows to add new interaction
processes relatively easily. In its first version~\citep{Ciardi2001}, the
code follows hydrogen ionization by point sources, taking into account
radiation emitted during recombinations. The physics of helium, a treatment
of the gas temperature and the possibility to include diffuse background
radiation were introduced later~\citep{Maselli2003}, as well as an improved
multi-frequency scheme~\citep{Maselli2009}. The latter version is referred to
as \crash~2\,. In \cite{Graziani2013} \crash~3 was introduced, featuring the
physics of atomic metals, several improvements necessary to account for
different source populations in reionization simulations
(\citealt{Kakiichi2017, Eide2018a}), optional coupling with hybrid pipelines
of galaxy formation \citep{Graziani2015, Graziani2017} and the capability to
propagate photons through multi-scale AMR grids \citep{Hariharan2017}. The
newest version, \crash~4 \citep{Graziani2018}, self-consistently treats
ionization by high-energy photons (up to $h\nu = \SI{10}{keV}$) as well as
secondary ionization by collisions with high-energy photo-electrons. The name
\crash\ will hereafter be used to refer to this version. Most of the features
mentioned in this paragraph can easily be turned on or off for a particular
simulation, depending on the problem at hand.

We will now briefly describe \crash\ by showing its workflow on the basis of
an example simulation that does not include metals or dust and does not
feature background radiation. The changes made to this workflow to account
for dust are discussed afterwards. The interested reader is referred to the
literature cited above for more details.

The starting point of a \crash\ simulation is a set of initial conditions
(ICs) provided on a three-dimensional Cartesian grid of $N_\mathrm{c}^3$
cells and a given linear size $L_\mathrm{b}$\,. The basic configuration
requires:
\begin{enumerate}%[label=\roman*)]
\item number density of the gas
  $n_\mathrm{g}\,\left[\mathrm{cm}^{-3}\right]$, as well as hydrogen and
  helium number fractions;
\item gas temperature $T\,\left[\mathrm{K}\right]$ and ionization fractions
  $x_{\HII}$, $x_{\HeII}$, $x_{\HeIII}$ at the
  initial time $t_0$;
\item coordinates, emission rates of ionizing photons ($\dot{N_j}$) and
  spectral energy distributions (SEDs) of all point sources $j$;
\item the simulation duration $t_\mathrm{s}$ and its starting redshift.
\end{enumerate}
ICs can either be manually created to perform tests, or they can be obtained
from N-body or hydrodynamics simulation snapshots providing realistic
configurations at fixed times. RT simulations are then performed in
post-processing by propagating photons through a single snapshot, or, e.g. in
the case of reionization simulations, by running \crash\ successively on a
set of snapshots, propagating the generated outputs from one to the next
(e.g.~\citealt{Ciardi2003, Ciardi2012, Graziani2015, Eide2018a}). During a
single RT run gas dynamics are not followed, but the density evolution of the
gas due to cosmological expansion as well as the cosmological redshift of the
propagated photons can be taken into account.

Once a set of ICs is loaded, \crash\ starts looping over the point sources,
emitting and propagating a photon packet from every source, until the
specified number of packets per source $N_\mathrm{p}$ is reached. The photon
count of each packet is given by the source emissivity and the simulation
time:
\begin{equation}
  N_j = \frac{\dot{N_j} t_\mathrm{s}}{N_\mathrm{p}}\,.
\end{equation}
These photons are distributed over user specified frequency bins in
accordance with the SED of the source $j$. The emission direction is
determined by randomly sampling the corresponding angular probability
distribution function. Each packet is then moved along the ray specified by
its direction and the source position. For every cell it crosses, the number
of photons $N_\nu$ in a frequency bin around $\nu$ is reduced by
\begin{equation}
  \label{eq:photon-absorption-in-cell}
  \Delta N_\nu = N_\nu (1- \mathrm{e}^{-\tau_\nu})\ ,
\end{equation}
where $\tau_\nu = \sum_i \tau_{i,\nu}$ with
$i\in \left\{\mathrm{\HI},\HeI,\HeII\right\}$ is the total optical depth of
the cell at frequency $\nu$. The partial optical depths are given by
\begin{equation}
\tau_{i,\nu} = n_i\sigma_i l,
\end{equation}
where $n_i$ and $\sigma_i(\nu)$ are, respectively, number density and
absorption cross section of species $i$, and $l$ is the geometrical path
casted by the ray through the cell. The absorbed photons are distributed
among the different species in the cell according to\footnote{This has been
  changed since~\cite{Maselli2003}; see
  also~\citet[Appendix~D]{Friedrich2012}.}
\begin{equation}
  \label{eq:ionization-number-in-cell}
  \Delta N_{i,\nu} = \frac{\tau_{i,\nu}}{\tau_\nu} \Delta N_\nu\,.
\end{equation}
This is used to compute ionization fractions and gas temperature (see
original papers for more details).

Finally, the photon packet is moved to the next cell crossed by its ray and
the process is repeated until the packet is eventually depleted or it leaves
the computational domain (although periodic boundary conditions can be
enabled). Photon packets created by recombination processes are emitted from
the corresponding cells when the emission criteria specified
in~\cite{Maselli2003} are met and are treated in the same way as the photon
packets emitted by a point source.

Finally note that, while \cite{Pierleoni2009} introduced an implementation of
the code treating the propagation of Ly\textalpha\ photons self consistently
with the ionizing continuum (by including both absorption and scattering), the
code version adopted here does not feature any line scattering physics, so that
it is not applicable to high density plasmas. Also note that in this work the
atomic metal functionality of \crash\ is not coupled with the dust module (see
\S~\ref{sec:silic-graph-pah}).

\section{Dust models}
\label{sec:dust-models}
Studying radiative transfer through dust and assessing its interplay with
ionized gas requires detailed knowledge of several dust properties, for
example optical coefficients, charging yields and sublimation temperatures.
Much remains unknown about cosmic dust in general, and there is, as of yet,
no consensus in the community on the composition and nature of dust even in
the galactic ISM. Consequently, many different models exist (e.g.
\citealt{Li1997, Zubko2004, Jones2013}). For this reason a numerical
implementation of dust in a radiative transfer code should be
model-independent (see~\S~\ref{sec:crash-with-dust}).

Moreover, there is no a priori reason to assume dust at high redshift to be
like the dust we observe in the local universe; in fact, evidence to the
contrary has been presented (e.g. \citealt{Kulkarni2016}). Likewise, one
should expect grain properties to vary depending on their environment
(cf. Fig.~\ref{fig:dustDiffISM}). In the context of this work, we will only
briefly investigate the processing of dust in \HII\ regions.

This section is organized as follows. We first present the dust properties we
require in \S~\ref{sec:dust-optProps} and then introduce the
Silicate-Graphite-PAH model in
\S~\ref{sec:silic-graph-pah}. \S~\ref{sec:dust-PAH}, finally, discusses the
contribution of PAHs to the dust optical depth.

\subsection{Optical properties}
\label{sec:dust-optProps}
Photon absorption and scattering by a dust grain are characterised by its
frequency-dependent absorption and scattering cross sections,
$\sigma_\mathrm{a}(\nu)$ and $\sigma_\mathrm{s}(\nu)$ respectively.
$\sigma_\mathrm{e} = \sigma_\mathrm{a} + \sigma_\mathrm{s}$ is referred to as
the extinction cross section, and the relative importance of scattering
against extinction is quantified by the albedo
$\widetilde{\omega} = \sigma_\mathrm{s} / \sigma_\mathrm{e}$. The scattering
asymmetry parameter $\left<\cos\theta\right>$, where $\theta$ is the
scattering angle and the brackets indicate averaging over scattering events,
complements the information about the capability of dust grains to deviate
incoming light. We customarily refer to the above-mentioned quantities as the
\textit{optical properties} of dust (see \citealt{Henning2010} for a recent
review or \citealt{Bohren1983} for an extensive treatment). They generally
depend on a series of grain properties (e.g. chemical composition,
solid-state structure, morphology) and environment-dependent conditions
(e.g. dust temperature, charge state) and can be determined both
theoretically (e.g. \citealt{Draine2003b, Draine2003c}) and experimentally
(e.g. \citealt{Henning2010} or see the \textit{Amsterdam-Granada light
  scattering database}\footnote{\url{http://www.iaa.es/scattering/}}).

Typically, the optical properties of cosmic dust can observationally only be
constrained at photon energies $h\nu < \SI{13.6}{eV}$ (e.g.
\citealt{Cardelli1989}), where the gas component of the ISM does not dominate
absorption, and at energies $\gtrsim \SI{e2}{eV}$, by means of X-ray
scattering haloes \citep[e.g.][]{Smith1998, Draine2003f}. In the ionizing-UV
band \SIrange{13.6}{200}{eV}, which is of central importance to the
ionization of the hydrogen and helium gas components, one generally has to
fully rely on predictions from theoretical models supported, when available,
by laboratory experiments.

\subsection{The Silicate-Graphite-PAH model}
\label{sec:silic-graph-pah}
We adopt optical properties provided by the so-called Silicate-Graphite-PAH
model (\citealt{Weingartner2001a} and \citealt{Li2001}; hereafter WD01 and
LD01 respectively). In this subsection we summarize the main assumptions of
the model and present some of its results, as they are necessary to
understand the implementation in \crash . We refer the interested reader to
the original papers for more details.

WD01 and LD01 assume a dust composition consisting of two chemically distinct
grain populations: one based on silicate and one based on carbon. For each
population, they assume a grain size distribution described by a modified
power-law with a smooth cut-off at large sizes
($a \sim \text{\SIrange{1}{10}{\micro\meter}}$) and a variable slope towards
the small-size cut-off at \SI{3.5}{\angstrom} (see eqs. 4--6 in WD01). Silicate
grains and large carbon grains (with an effective grain radius
$a\gtrsim \SI{100}{\angstrom}$; see \citealt[\S~2.1]{Weingartner2001b}) are
assumed to be spherical and composed of olivine and graphite respectively;
their optical properties are computed in the framework of Mie theory
(e.g. \citealt{Bohren1983}). For the carbonaceous population two log-normal
distributions peaking at \SI{3.5}{\angstrom} and \SI{30}{\angstrom} are added
to provide additional small grains. Small carbon grains
($a \lesssim \SI{20}{\angstrom}$) are given the optical properties of
polycyclic aromatic hydrocarbons (PAHs)\footnote{ The properties used are
  guided by laboratory measurements and therefore the actual geometry of the
  PAHs is irrelevant for this purpose. When modelling aspects where this is not
  true, small PAHs are usually assumed to be planar while larger ones are
  assumed to be spherical \citep[e.g.][]{Hoang2010}.}, and transitional
properties are used for intermediate size carbon grains (see also
Eq.~\eqref{eq:tau-d}). WD01 and LD01 tune their size distributions in order for
their dust model to reproduce the observed dust extinction and infrared
emission of the diffuse Milky Way ISM. The \cite{Fitzpatrick1999}
$R_V$-parametrization of the extinction curve with $R_V=3.1$ is chosen to
represent the observed extinction, where $R_V=A_V/E(B-V)$ is the ratio of
visual extinction to reddening. The model can also reproduce extinction along
many other lines of sight through the ISM of the Milky Way and also of the
Magellanic Clouds, where typical extinction differs from that of our galaxy
(see \S~3.2 of WD01).

\begin{figure}
  \centering
  \includegraphics[width=\linewidth]{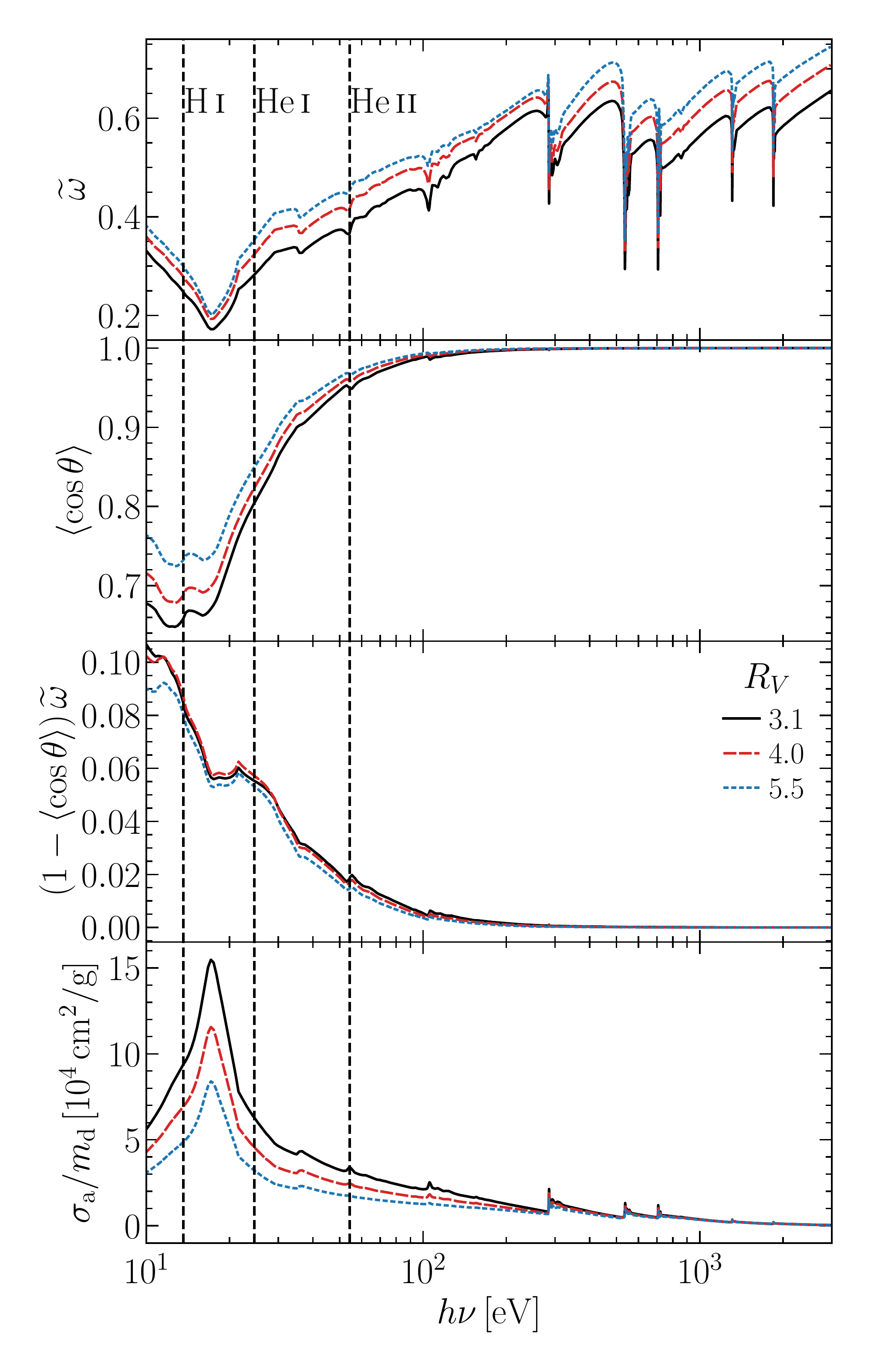}
  \caption{From top to bottom the panels show as functions of photon energy:
    the albedo $\widetilde{\omega}$, the scattering asymmetry parameter
    $\left<\cos \theta \right>$,
    $\left(\num{1} - \left<\cos \theta \right>\right) \widetilde{\omega}$ and
    the absorption cross section per dust mass, $\sigma_{\rm a}/m_{\rm
      d}$. The different lines correspond to dust mixtures that reproduce
    extinction along Milky Way lines of sight with $R_V$ values of \num{3.1}
    (solid black), \num{4.0} (dashed red) and \num{5.5} (dotted blue)
    \protect\citep{Weingartner2001a, Li2001,Draine2003a, Draine2003b,
      Draine2003c}. The ionization potentials of hydrogen and helium are
    marked with vertical dashed lines. See text for more details.}
  \label{fig:dustDiffISM}
\end{figure}

A later version of the model \citep{Draine2003a, Draine2003b, Draine2003c}
provides corrected dust abundances as well as updated material constants. The
optical properties of the dust mixture are computed in a wide energy range
($\sim \text{\SIrange{e-4}{e4}{eV}}$), using anomalous diffraction theory
\citep{Hulst1957} instead of Mie theory at X-ray energies. These new
data\footnote{New and old data are available online at
  \url{www.astro.princeton.edu/~draine/dust/dustmix.html}\,.} are plotted as
functions of photon energy in Fig.~\ref{fig:dustDiffISM} for
$R_V = \num{3.1},\ \num{4.0}$ and \num{5.5}\,. More specifically, we adopt
the WD01 case~A size distributions with carbon abundance per H atom in the
log-normal populations $b_\mathrm{C}=\SI{60}{ppm}$ and total carbon dust
abundance corrected by a factor \num{0.93} for $R_V=3.1$, with
$b_\mathrm{C}=\SI{40}{ppm}$ corrected by \num{1.18} for $R_V=4.0$ and with
$b_\mathrm{C}=\SI{30}{ppm}$ corrected by \num{1.42} for $R_V=5.5$ (see WD01
and \citealt{Draine2003a}). While the three lines differ, their qualitative
behaviour is very similar, and thus we will restrict ourselves to discussing
the $R_V = 3.1$ case. We also restrict the discussion to H-ionizing energies
($h\nu \geq \SI{13.6}{eV}$) as \crash\ performs RT only in this photon energy
range. Note, additionally, that more recent cross sections and material
constants, taking into account new observations and laboratory data, have
been published for the carbonaceous model-grains \citep{Draine2007,
  Draine2016}. These latest updates are not taken into account here, since
the older optical data plotted in Fig.~\ref{fig:dustDiffISM} are readily
available and easily sufficient for the purposes of this work.

The albedo (top panel) starts out at a value of $\sim 0.2$ around
\SI{13.6}{eV} and remains below $\sim\num{0.5}$ in the entire UV regime
($\SI{13.6}{eV} < h\nu \lesssim \SI{200}{eV}$), i.e. dust absorption
dominates scattering at these energies. Entering the soft X-ray domain
($\SI{0.2}{keV} < h\nu < \SI{3}{keV}$), it rises approximately linearly with
$\log (h\nu)$ to $\sim \num{0.6}$ at \SI{300}{eV}. From this point onwards up
to \SI{3}{keV} it remains roughly constant on average, so that at these
energies more than \SI{50}{\percent} of the photon-dust interactions are
scattering events. The scattering asymmetry parameter (second panel) rises
quickly from $\approx\num{0.7}$ at \SI{13.6}{eV} to unity at \SI{100}{eV},
i.e. in our energy regime the dust grains are mostly to purely forward
scattering.

As \crash\ currently neglects scattering in its ray tracing
algorithm\footnote{We repeat here that the implementation of radiation
  scattering in a code accounting for gas ionization and temperature as well
  as dust as function of time, would severely increase the computational
  requirements of a single run or even impede reionization
  simulations. Sophisticated implementations of radiation scattering on dust
  grains exist (e.g. \citealt{Camps2015, Gordon2017}) but they are generally
  limited to time independent algorithms and do not account for the gas
  component.}, we investigate the impact of this approximation using
$\left(\num{1}-\left<\cos\theta\right>\right) \widetilde{\omega}$ (third
panel). This parameter measures how strongly photons are deviated from their
initial direction of propagation upon interacting with a grain. If it is
zero, no deviation takes place and scattering can safely be neglected. In the
range \SI{13.6}{eV}--\SI{3}{keV}, it remains below \num{0.1}, indicating that
neglecting scattering is an acceptable approximation in our ray-tracing
scheme, at least with the present dust model and in applications to diffuse
gas, as those of our interest. This scheme is, on the other hand, not
appropriate to model protoplanetary discs, for instance, and using a
different dust model might invalidate our argument. We plan to address
scattering in a future release after implementing other dust processes that
directly impact grain charging and gas ionization.

The bottom panel of Fig.~\ref{fig:dustDiffISM}, finally, shows the absorption
cross section per dust mass. It peaks at $\sim\SI{17}{eV}$ (i.e. between the
\HI\ and \HeI\ ionization potentials) and progressively decreases with
increasing energy, showing some structure owed to the inner shells of grain
constituents.

\begin{figure}
  \centering
  \includegraphics[width=\linewidth]{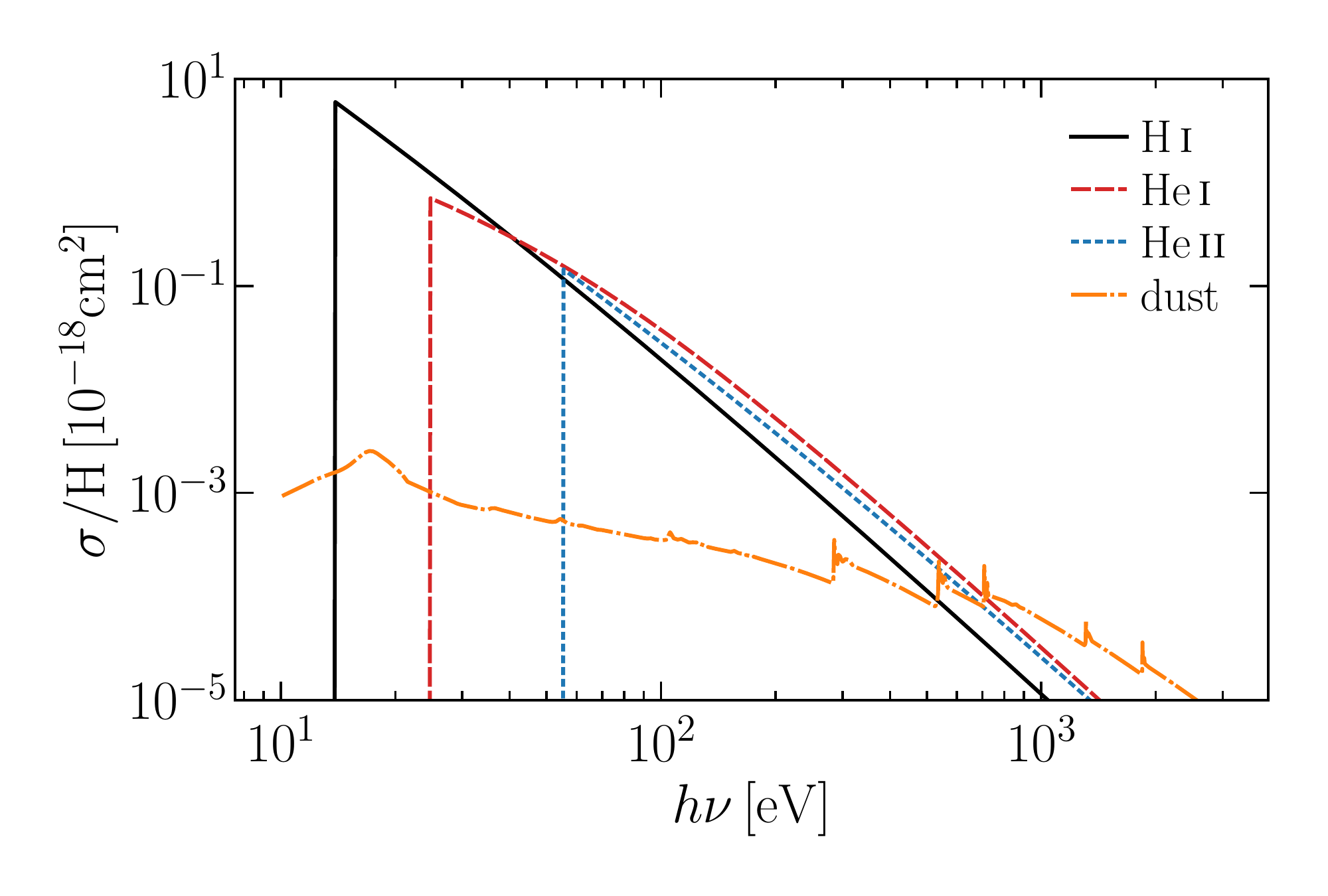}
  \caption{Ionization cross section per hydrogen nucleus as a function of
    energy for neutral hydrogen (solid black line), neutral helium (dashed
    red) and singly ionized helium (dotted blue)
    \protect\citep{Verner1996}. Also shown is the absorption cross section of
    dust per hydrogen nucleus for $R_V=\num{3.1}$ (dash-dotted
    orange). Protosolar gas composition and a GDR of 124 were assumed. See
    text for details.}
  \label{fig:gasVSdust}
\end{figure}

Since the main goal of our implementation is to investigate the RT through dust
and gas (H,~He), it is important to compare the relevant cross sections at
fixed photon energy. Fig.~\ref{fig:gasVSdust} shows the dust absorption cross
section (dash-dotted orange) together with the ionization cross sections of
neutral hydrogen (solid black) and neutral helium (dashed red)
\citep{Verner1996}; these are the primary species competing with dust grains in
absorbing ionizing photons. The cross sections are normalized per H nucleus. We
assume protosolar gas composition (\citealt[Table 1.4]{Draine2011}; see also
\citealt{Asplund2009}) and a gas-to-dust mass ratio (GDR) of \num{124}, as
suggested by WD01 and LD01 for the diffuse ISM of our galaxy. With the above
assumptions, it is evident that in a neutral gas-dust mixture, dust will only
be relevant to absorption at energies of several hundred \si{eV}, but note that
the GDR can vary strongly from $\sim \num{10}$ to $\sim \num{e6}$
(e.g. \citealt{Vuong2003, Draine2007a, Sandstrom2013, Remy-Ruyer2014,
  Liseau2015, Owen2015, Mancini2016}), depending on the galactic environment.
Unlike gas (which becomes ionized and thus transparent), however, the opacity
of dust in photo-ionized regions does not change significantly in time unless
the UV field is strong enough to destroy the grains\footnote{It should be noted
  that this is not generally true. In molecular regions (which we do not study
  here), e.g., photo-detachment from negatively charged grains can appreciably
  change the grain cross section \citep[e.g.][]{Weingartner2001b}. Also note
  that grain destruction is not yet implemented in \crash\ (see
  \S~\ref{sec:crash-with-dust}) and that different grain types (both size and
  composition) are expected to behave very differently when exposed to
  radiation.}. Therefore dust can provide time-persistent opacity over a wide
photon energy range and might be an important contributor to absorption also at
lower energies. For example, a typical stellar radiation field can easily
ionize both \HI\ and \HeI\ so that the dust grains will, after a certain point
in time, compete only with singly ionized helium (dotted blue) while
continuously absorbing energy from the UV field in \SIrange{13.6}{54.4}{eV}. In
reality, however, the situation is much more complex because (i) the presence
of atomic metals in the gas mixture \citep[for a discussion including abundance
constraints on dust models see][]{Draine2003a, Draine2011} alters the total
optical depth and (ii) the ionization fractions and recombination rates of each
species are complex functions of local conditions. It is thus not possible to
make a general quantitative statement on the importance of dust absorption in
ionized gas without detailed numerical modeling. We defer a more detailed
discussion and the inclusion of gas phase metals to future investigations.

Our choice of the Silicate-Graphite-PAH model for the purposes of this work
is owed to the fact that it is widely used, easily accessible and
comparatively simple. On the other hand, this model is primarily designed to
explain the various spectral features of dust extinction curves and dust IR
emission on an observation by observation basis, i.e. it offers no
straightforward formalism to model the evolution of a given dust
population. For example, the PAH contribution to the total absorption cross
section can not easily be isolated. As PAHs are susceptible to
photo-dissociation, it is important to quantify the error introduced by
assuming their continuous presence during the RT process. This point is
addressed in the following section.

\subsection{Contribution of PAHs to the dust optical depth}
\label{sec:dust-PAH}

Dust mixtures are known to be sensitive to both UV radiation and gas phase
transitions from neutral to ionized. The PAH component, in particular, is
susceptible to restructuring and destruction via photo-dissociation
\citep[e.g.][]{Voit1992, Jochims1994, Page2001} and collisions with free
electrons and ions present in a hot, ionized gas
\citep[e.g.][]{Micelotta2010, Micelotta2010a, Bocchio2012}. The efficiency of
the above processes, however, as well as the change in PAH composition
induced by them, are not fully understood and are subject of thorough
investigations \citep[e.g.][]{Zhen2016}. Observations, nevertheless, suggest
PAHs to be systematically depleted in galaxies hosting an active galactic
nucleus (AGN) \citep{Roche1991, Smith2007, Jensen2017} or in \HII\ regions
(e.g. \citealt{Kassis2006, Whelan2013, Stephens2014, Salgado2016}; Chastenet
et al., in prep.), so that they seem to provide conditions conducive to PAH
destruction \citep[but see][]{Compiegne2007}.

While many processes of PAH evolution have been implemented for circumstellar
disks \citep{Visser2007, Siebenmorgen2010} or to model specific features of
galactic \HII\ regions \citep{Giard1994}, a consistent time dependent RT
approach including dust mixtures and gas is still missing. It is, in fact,
not clear how current models for dust mixtures (see section above) should be
modified to derive cross sections excluding the photo-dissociated PAH
component.

In this work we perform a first step in this direction by recomputing the
cross section of a dust mixture without the PAH component in the framework of
the Silicate-Graphite-PAH model. The result will be used in \S~\ref{sub:
  PAH-HII} to asses the impact of the PAHs on the size of ideal dusty \HII\
regions and in \S~\ref{sec:rt-dusty-web} with a more complex dust
distribution and source configuration.

For an immediate comparison with the discussion of the previous section, here
we discuss the resulting absorption cross sections, while the details of the
modelling are deferred to Appendix~\ref{sec:nopah-dust-model}.

\begin{figure}
  \centering
  \includegraphics[width=\linewidth]{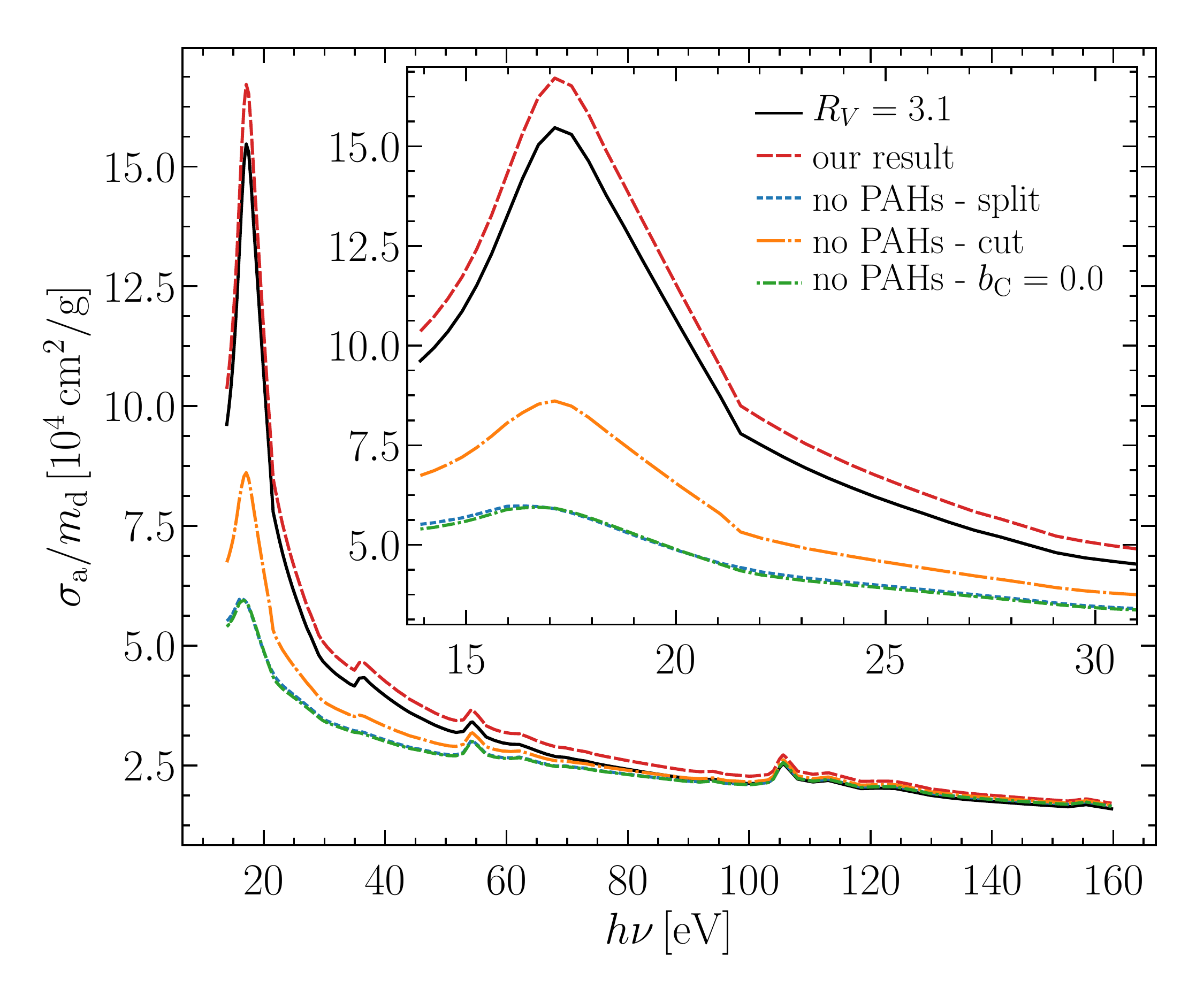}
  \caption{Dust absorption cross section per mass as a function of photon
    energy for the Silicate-Graphite-PAH model ($R_V=\num{3.1}$) as in
    Fig.~\ref{fig:dustDiffISM} (solid black) and recomputed by the authors
    as detailed in Appendix~\ref{sec:nopah-dust-model} (solid red). Cross
    sections obtained excluding the PAH component are shown for three
    different assumptions in removing PAHs: the dotted blue line is the
    result of splitting the carbon grain size distribution into a graphite
    and a PAH component, the dash-dotted orange line is obtained by cutting
    the size distribution at a grain radius corresponding to \num{1000}~C
    atoms, and finally for the dash-dash-dotted green line we set
    $b_\mathrm{C}=\SI{0.0}{ppm}$, thus removing the two log-normal grain
    size distributions. For details on these three approaches see
    Appendix~\ref{sec:nopah-dust-model}. A zoom-in view of the curve in the
    most relevant energy range \SIrange{13.6}{30}{eV} is provided by the
    inset.}
  \label{fig:PAH-dest}
\end{figure}

Fig.~\ref{fig:PAH-dest} shows the dust absorption cross sections (per mass of
full dust mixture) as a function of photon energy. Before introducing any
changes in the dust composition, we note that the original
Silicate-Graphite-PAH model with $R_V=\num{3.1}$ (solid black line) and our
own result when PAHs are included (dashed red line) are in good agreement,
with a maximum difference of \SI{9}{\%} at \SI{21.5}{eV}\footnote{The
  systematic offset between cross sections is likely due to the fact that we
  adopted a slightly different version of the Wiscombe-Mie code (Warren
  J. Wiscombe, Mie Scattering Calculations: Advances in Technique and Fast,
  Vector-Speed Computer Codes, NCAR/TN-140+STR, NCAR TECHNICAL NOTE Jun); see
  also Appendix~\ref{sec:nopah-dust-model}.}.

A significant reduction of the cross section by a factor of two (orange line,
cut model) or three (blue and green lines for split and
$b_\mathrm{C}=\SI{0.0}{ppm}$ models, respectively) relative to the original
data is found at its peak ($\sim\SI{17}{eV}$) and it remains considerable
($\sim\SI{20}{\%}$) in the energy range up to \SI{30}{eV}. The differences
progressively decrease to \SI{5}{\%} at $\sim\SI{60}{eV}$, and remain at or
below this value at higher frequencies because all size distributions
converge to the original model, i.e. the PAH component does not contribute
significantly anymore to the mixture cross section.

In summary, the PAH component provides an important contribution in
absorption in the photon energy range \SIrange{13.6}{60}{eV}, while its
presence is negligible in the remaining part of the spectrum. In
\S~\ref{sec:results} we investigate the impact of this on gas ionization and
temperature, using the $b_\mathrm{C}=\SI{0.0}{ppm}$ approach to obtain an
upper limit on the impact.

\section{Numerical implementation of dust in \crash}
\label{sec:crash-with-dust}
Here we describe the modifications which enable \crash\ to account for dust
absorption in RT simulations. We start discussing how we abstracted this
problem to achieve an implementation that is extendable and independent of
any specific dust model.

The most general way to obtain the necessary data from a dust model is by
interpolation of pre-computed tables, which is therefore our method of
choice. Any chosen dust model might provide properties for a predefined mix
of dust species (as the Milky Way model introduced earlier), or it might
provide properties for individual dust species; it is therefore clear that a
general dust RT code should support the concept of dust species. Furthermore,
in realistic applications dust will not be distributed homogeneously in
space, and its properties (composition, ionization, grain size, temperature,
etc.) may also be functions of location and time. An RT code accounting for
dust, therefore, needs to be able to track not only the dust abundance but
also its properties, possibly per species, at each point of the discretised
domain during the entire RT process. This has the additional advantage that
radiation effects on dust grains (such as heating, ionization, metamorphism
and destruction) can be accounted for in detail, and their coupling to gas
temperature and ionization can be modelled in the photo-ionization
equations. The initial dust distribution and properties can then be
self-consistently modified during the RT.

While the currently implemented dust treatment is very simple in that it only
accounts for dust absorption (i.e. removal of photons from the ionizing flux
by dust), the above considerations were kept in mind during the development,
so that introducing a more refined dust modelling is relatively easy in the
present code framework.

We created a new software module to include dust following the standard
\crash\ framework: the dust module is configurable by an ASCII text file,
which is loaded along the already existing configuration files for other
modules. The file specifies what dust species are considered and from where
to load the tables with their optical properties and other dust-related
ICs. These are appended to the list presented in \S~\ref{sec:crash}:
\begin{enumerate}%[label=\roman*)]
  \setcounter{enumi}{5}
\item dust mass density $\rho_\mathrm{d}\,$\,[g\,cm$^{-3}$]\,;
\item fraction of mass $y_k$ in each dust species $k$.
\end{enumerate}
To switch between different models, only a configuration change is required,
as model-specific values are loaded from files during simulation
initialization.

A FORTRAN \textsc{type} DUST\_SPECIES has been created to represent a single
dust species and to hold the tables provided by the adopted model. The 3D
variability of dust in the computational domain is described with a map of
DUST\_IN\_CELL \textsc{type} variables holding the mass density of dust
$\rho_\mathrm{d}$, the species mass fractions $y_k$ and the absorption cross
section per mass $\diff\sigma_\mathrm{a}(\nu)/\diff m_\mathrm{d}$ in a single
cell. According to the frequency sampling adopted in the \crash\ spectra, the
absorption cross section is computed at each centre-bin frequency $\nu$ by
taking the mass fraction weighted average of the species cross sections:
\begin{equation}
  \tder{\sigma_\mathrm{a}(\nu)}{m_\mathrm{d}}
  = \sum_k y_k \tder{\sigma_{\mathrm{a},k}(\nu)}{m_{\mathrm{d},k}}\,.
\end{equation}
These can be obtained by interpolation of the DUST\_SPECIES tables during
\crash\ initialization to avoid impacting algorithm performance during photon
propagation.

When a photon packet crosses a cell, we account for dust absorption by adding
a new term to the optical depth $\tau_\nu$ from
equation~(\ref{eq:photon-absorption-in-cell}):
\begin{equation}
  \tau_\nu' = \tau_\nu
  + \rho_\mathrm{d}\tder{\sigma_\mathrm{a}(\nu)}{m_\mathrm{d}} l\,,
\end{equation}
with the dust mass density $\rho_\mathrm{d}$ in the cell.
Equation~(\ref{eq:ionization-number-in-cell}) thus becomes:
\begin{equation}
  \Delta N_{i,\nu} = \frac{\tau_{i,\nu}}{\tau'_\nu} \Delta N_\nu\,,
\end{equation}
where $i$, as before, labels the gas components. The photons absorbed by dust
are simply removed from the flux; they do not affect the dust status in any
way, i.e. its contribution to the optical depth of the cell stays constant in
time. This is, naturally, a point to be improved upon in the future by taking
into account radiation processes that modify dust properties and by including
the gas to dust coupling they introduce. Photoemission of electrons from
grains, for example, is already being implemented and will be included in a
future version.

Apart from the modifications discussed here, the RT proceeds as in
\S~\ref{sec:crash}; specifically the gas ionization equations have not been
modified.

\section{Results}
\label{sec:results}
In this section we present the results of RT simulations in different
idealized environments: a dusty \HII\ region created by an O--type star
(\S~\ref{sec:stroemgren-sphere}) and a small, cubic cosmological volume with
an edge length of $\num{0.5}h^{-1}\,\mathrm{cMpc}$ in which a few bright
stellar type sources propagate their ionizing radiation through dust-enriched
gas. The latter test case (\S~\ref{sec:rt-dusty-web}) is an adapted version
of Test~4 introduced in \cite{Iliev2006} and further discussed in
\cite{Graziani2013}.

The purpose of these simulations is to verify that our code produces sensible
results in simplified configurations mimicking different astrophysical
environments in which dust is directly observed (e.g. galactic \HII\ regions)
or its existence is inferred (e.g. high-$z$ cosmic webs).

Hereafter, the gas is always assumed to be composed of \SI{92}{\percent} H
and \SI{8}{\percent} He by number. At the beginning of any simulation the gas
is assumed to be fully neutral and to have a temperature of \SI{100}{K}. In
simulations featuring dust, we assume the dust composition to be the same
throughout the entire simulated volume and time.

\subsection{Dusty, stellar-type \HII\ regions}

\label{sec:stroemgren-sphere}

The configuration we study here oversimplifies the typically clumpy,
chemically diverse, dynamic and turbulent star forming environment, by making
the following assumptions\footnote{It should be noted that the problem of the
  expansion of dusty \HII\ regions is far more complicated as dust
  efficiently couples with radiation, correlating the hydro-dynamics with the
  RT \citep[e.g.][]{Faucher-Giguere2012, Paladini2012, Akimkin2015,
    Akimkin2017}. Self-consistent modelling of this problem is beyond the
  scope of the current work.}:

\begin{itemize}

\item The cubic simulation volume of $\left(\SI{85}{pc}\right)^3$ contains
  homogeneous and static gas with a number density
  $n_\mathrm{gas}=\SI{1}{cm^{-3}}$. This value is typical of the diffuse
  ionized medium and should be appropriate for an evolved, low-density \HII\
  region (e.g.~\citealt{Anantharamaiah1986, Roshi2000}). The Cartesian grid
  mapping the volume has a resolution of $N_\mathrm{c} = \num{256}$ cells per
  side, corresponding to a spatial resolution of roughly \SI{0.3}{pc}.

\item We pollute the medium with dust using a fixed GDR. Our reference value
  is $\text{GDR}=\num{124}$, as proposed by WD01 and LD01 for the diffuse
  ISM. We also explore cases with $\text{GDR}=\num{50}$ and
  \num{1000}. Similarly, we adopt $R_V=\num{3.1}$ in the reference case and
  explore models with $R_V=\num{4.0}$ and $\num{5.5}$
  (cf. Fig~\ref{fig:dustDiffISM}).

\item The O--type star is modelled as an ideal black-body
  source\footnote{While in \crash\ arbitrary spectra can be assigned to
    sources and using a realistic spectrum from a stellar atmosphere model is
    straightforward, here we opted against this choice as results with an
    idealized spectrum are of easier interpretation and because the main aim
    of this test case is to investigate the impact of dust and not to compare
    our results to observations.} with spectral temperature
  $T_\mathrm{bb}=\SI{4e4}{K}$ and emission rate of ionizing photons
  $\dot{N}=\SI{e49}{s^{-1}}$. These values are appropriate for a powerful
  O--star \citep{Martins2005}. The source spectrum spans the energy range
  (\num{13.6}--\num{160})\,eV and it is discretised into
  $N_\mathrm{f} = \num{67}$ frequency bins with adaptive spacing chosen such
  as to ensure good sampling of the relevant cross sections
  (cf. Fig.~\ref{fig:gasVSdust}). We use \num{2e9} photon packets to sample
  the source emission. Such a high number is necessary to reach convergence,
  since especially the dust free and the $\text{GDR}=1000$ cases proved to be
  very sensitive to Monte Carlo noise.

\item As the lifetime of a massive O--type star ($\lesssim\SI{10}{Myr}$,
  e.g. \citealt{Raiteri1996}) is typically much longer than the ideal
  ionization timescale for a Str\"omgren sphere
  \citep[Eq.~(15.6)]{Draine2011}, the \HII\ regions observed in our Galaxy
  are often assumed to be in equilibrium. We run our simulations for
  $t_\mathrm{f}=\SI{10}{Myr}$ and find that ionization equilibrium (in H and
  He) is reached after approximately \SI{1}{Myr}\footnote{To reach
    equilibrium, time dependent RT codes require several ideal ionization
    times ($\sim\SI{e5}{yr}$ in this case), as simple analytic estimates of
    such timescales do not account for recombinations that have to be
    balanced by the ionizing source \citep[e.g.][]{Iliev2006}.}, while
  temperature equilibrium is only reached towards the end of the simulations.
\end{itemize}

\begin{figure}
  \centering
  \includegraphics[width=\linewidth]{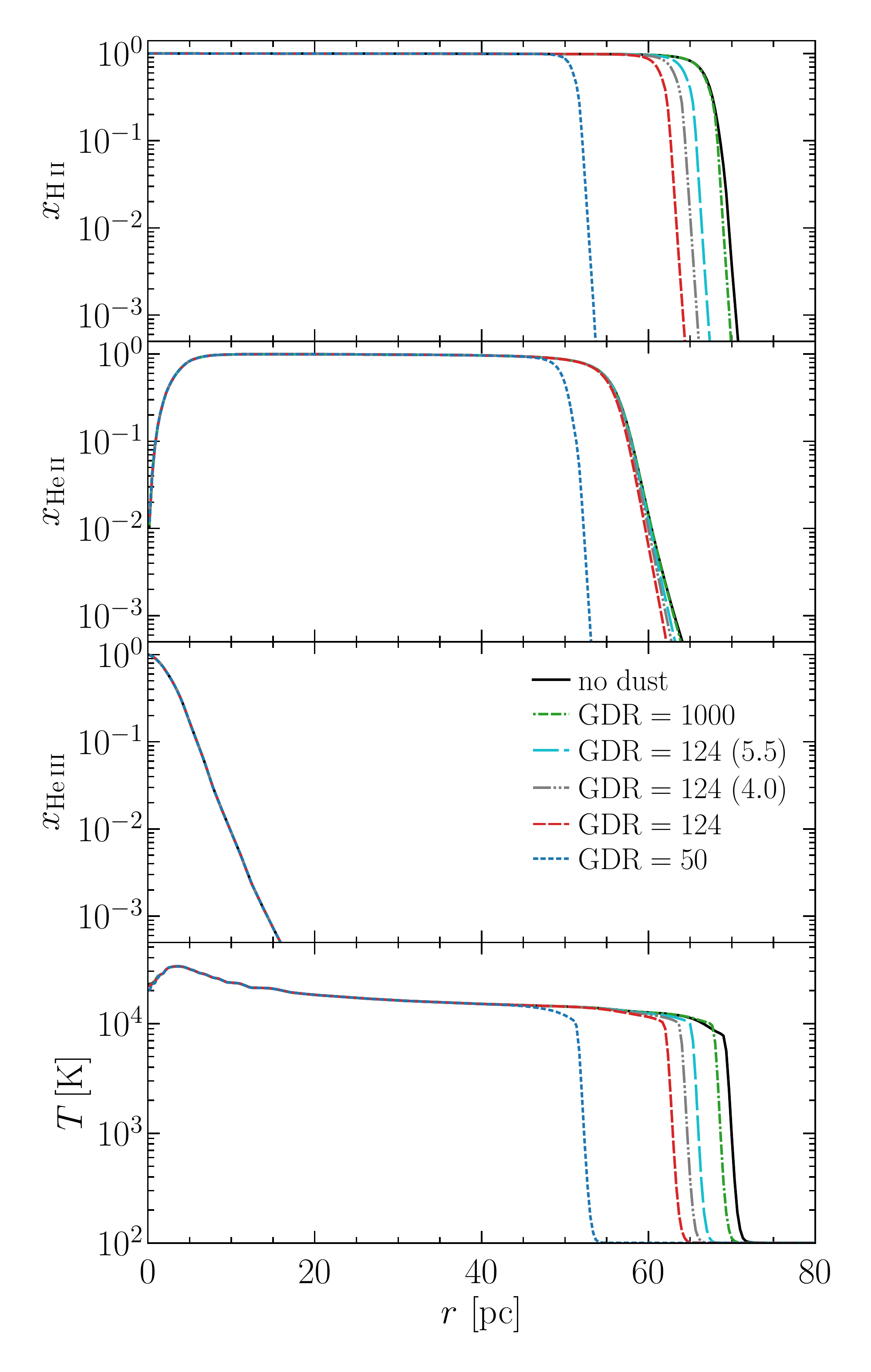}
  \caption{Spherical averages of (from top to bottom) \HII\ fraction, \HeII\
    fraction, \HeIII\ fraction and $T$ as functions of distance from the
    black-body source after \SI{10}{Myr} for $R_V=\num{3.1}$ and various GDRs
    as indicated in the labels. As a comparison, we also plot the dust free
    case (solid black line) and our reference GDR of \num{124} but with
    $R_V=\num{4.0}$ and \num{5.5} (indicated in brackets; see
    \S~\protect\ref{sec:silic-graph-pah}).}
  \label{fig:BBSphAvg}
\end{figure}

Fig.~\ref{fig:BBSphAvg} shows spherical averages of hydrogen and helium
ionization fractions as well as gas temperature as functions of distance $r$
from the stellar source. All lines show results at the final simulation time
$t_\mathrm{f}$, but correspond to different values of GDR and/or dust
models. For comparison we also show the dust-free case. Moreover, we define,
somewhat arbitrarily, the radius of the ionization front (I-front) as the one
at which $x_\HII$ drops below \num{0.9}\,.

First note that, in absence of dust, \crash\ computes an I-front position at
radius $r \sim \SI{64}{pc}$. When the medium is polluted by dust using our
reference values (dashed red line), the I-front recedes by $\sim \SI{4}{pc}$
(or \SI{6}{\percent}) due to the additional contribution to the total optical
depth. Also note that homogeneously distributed dust does not change the
slope of the ionization front, which maintains its sharp transition to
neutral gas. The increase in optical depth naturally depends on the value of
GDR and on the model assumptions leading to variations in the corresponding
cross section (see Fig.~\ref{fig:dustDiffISM}). Changing $R_V$ at constant
GDR, for example, produces small displacements consistent with the lowering
of the dust cross section with increasing $R_V$ value. In
Appendix~\ref{sec:comparison-petrosian} we compare our code to the analytic
solution for $x_\HII(r)$ by \cite{Petrosian1972}.

Close to the source helium is doubly ionized, and $x_\HeII$ is
correspondingly smaller than unity. Absorption by dust does not have a strong
effect on the helium ionization fractions inside the \HII\ region until the
GDR value is decreased below \num{100}, at which point also the \HeII\ front
begins to recede. This is due to the dust cross section being a factor of two
or three larger at \HI-ionizing energies than at \HeI-ionizing
energies\footnote{We find that when the \HII\ and \HeII\ fronts coincide in
  the dust free medium, as is the case for very hot black-body spectra
  \citep[cf.][Ch. 15]{Draine2011}, they recede jointly in dust polluted
  media.}.

The gas temperature profile (bottom panel) reaches values around
$T\sim \SI{e4}{K}$ where hydrogen is ionized and quickly drops in the neutral
region. Helium provides some additional heating close to the source. As in
the case of $x_\HII$, absorption by dust reduces the radius of the
heated bubble but does not change the structure in the inner region.

\begin{figure}
  \centering
  \includegraphics[width=\linewidth]{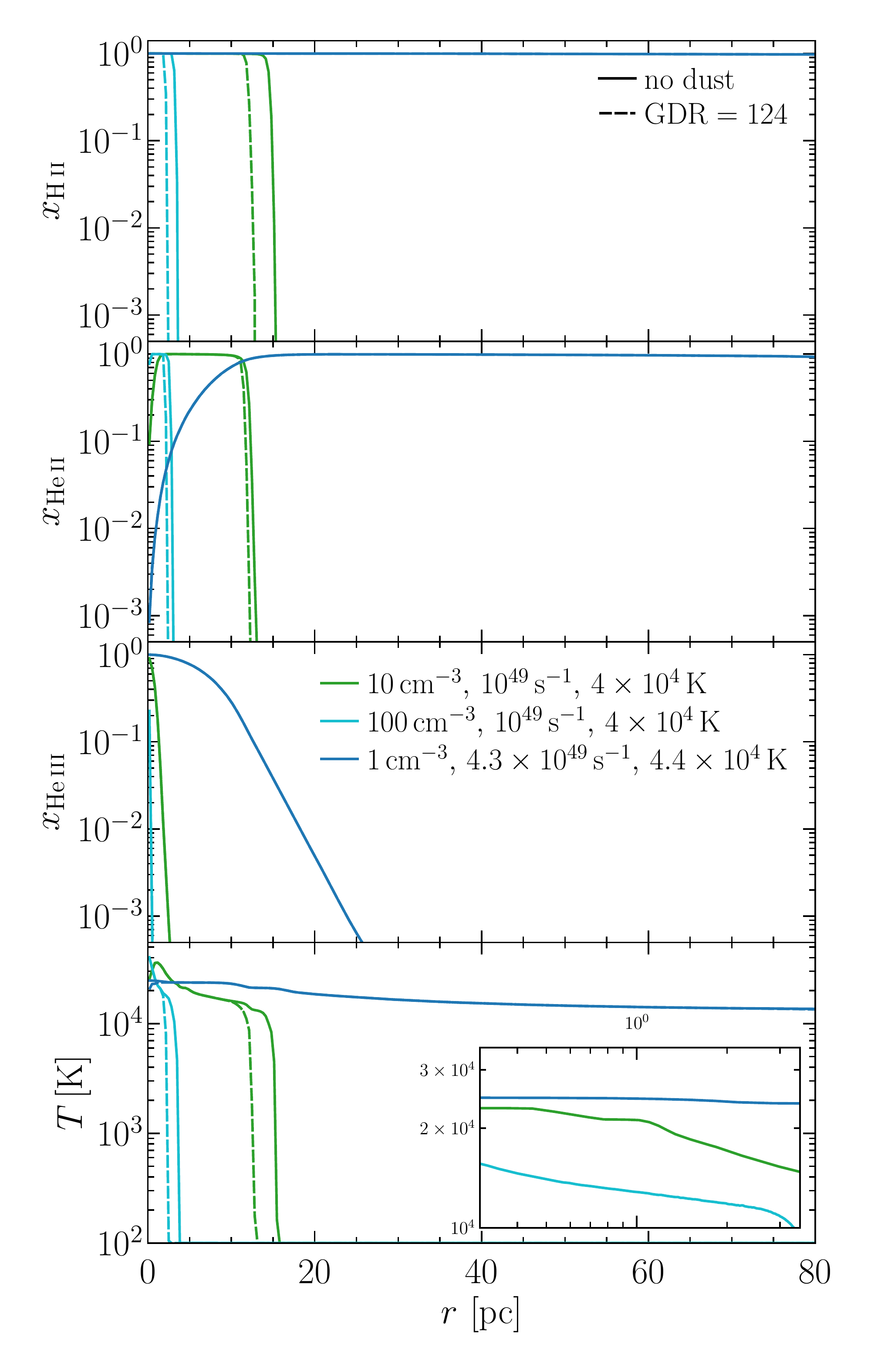}
  \caption{As Fig.~\protect\ref{fig:BBSphAvg} but showing results of
    simulations varying the gas number density and source ionization power:
    ($n_\mathrm{gas}$, $\dot{N}$, $T_\mathrm{bb}$) = (\SI{10}{cm^{-3}},
    \SI{e49}{s^{-1}}, \SI{4e4}{K}) in green, (\SI{100}{cm^{-3}},
    \SI{e49}{s^{-1}}, \SI{4e4}{K}) in light blue, and (\SI{1}{cm^{-3}},
    \SI{4.3e49}{s^{-1}}, \SI{4.4e4}{K}) in blue.  Dashed lines refer to
    simulations with $\text{GDR}=\num{124}$, while solid lines correspond to
    results for dust free media. In the bottom panel we show the temperature
    profile close to the source for high resolution runs; see text for
    details. A colour version of this figure is available in the online
    journal.}
  \label{fig:param-expl}
\end{figure}

In Fig.~\ref{fig:param-expl}, using the same layout as that of
Fig.~\ref{fig:BBSphAvg}, we show the results of exploring different values of
the gas number density $n_\mathrm{gas}$ and source ionizing power
(i.e. emissivity $\dot{N}$ and black-body temperature $T_\mathrm{bb}$), but
keeping $\text{GDR}=\num{124}$ (dashed lines). The values chosen for
$T_\mathrm{bb}$ and $\dot{N}$ are the highest mentioned in \citet[Table
1]{Martins2005} for solar metallicity O V stars. For each simulation we also
show the corresponding dust free results (solid lines). Our goal here is to
obtain a feeling for the possible variation in size of dusty \HII\ regions as
predicted by our code.

By increasing the gas number density to \SI{10}{cm^{-3}} and
\SI{100}{cm^{-3}}, the front recedes to \SI{12}{pc} and \SI{2}{pc},
respectively, compared to \SI{64}{pc} for our reference run. The helium
ionization regions show the expected behaviour, increasing in size with
decreasing density/increasing emissivity and vice versa. Note in the bottom
panel that the temperature close to the source is predicted to be higher for
weaker sources/higher gas density. This is clearly unphysical and it is owed
to the fact that for consistency we used the same spatial resolution of
\SI{0.3}{pc} for all simulations. When we choose more appropriate resolutions
to resolve the smaller Str\"omgren spheres produced at higher densities, this
problem is resolved (see inset in last panel).

The effect of dust absorption, again, consists in reducing the size of the
ionized regions, while leaving their inner structure unchanged.

We also briefly discuss the effect of dust on ionization fronts in media of
lower densities ($n_\mathrm{gas}\sim\SI{e-3}{cm^{-3}}$), such as the diffuse
IGM. Since photon mean free paths are relatively long, transitions from
ionized to neutral gas tend to be more extended and feature long
low-ionization tails in these cases. We find that the primary effect of dust
pollution then is the cut-off of these tails and that only low values of GDR
result in a size reduction of the highly ionized region. This effect can also
be seen in the simulations we discuss in \S~\ref{sec:rt-dusty-web}.

\cite{Hunt2009} and \cite{Draine_2011} compiled several observational data
sets of galactic and extra-galactic \HII\ regions. \cite{Hunt2009} interpret
the observed size-density relation in terms of dynamical semi-analytic models
including dust absorption and featuring time dependent $\dot{N}$. Starting
from different initial densities they account for \HII\ region expansion, but
retain the assumption of a homogeneous medium. \cite{Draine_2011}, on the
other hand, interprets the observations in terms of static models that
account for radiation pressure on dust creating central densities lower than
those close to the ionization front. Our model equilibrium \HII\ regions are
too small for a given density, even in dust free cases, to agree with the
observational results. This might indicate that our values for $\dot{N}$ are
too low \citep[cf.][\S~4.4]{Hunt2009}, or that the assumption of a
homogeneous medium should be removed by accounting for spatial variations of
gas density or clumpiness of dust polluted regions. A detailed comparison of
these models to our code could provide interesting insights but would go
beyond the scope of this paper.

\begin{figure}
  \centering
  \includegraphics[width=\linewidth]{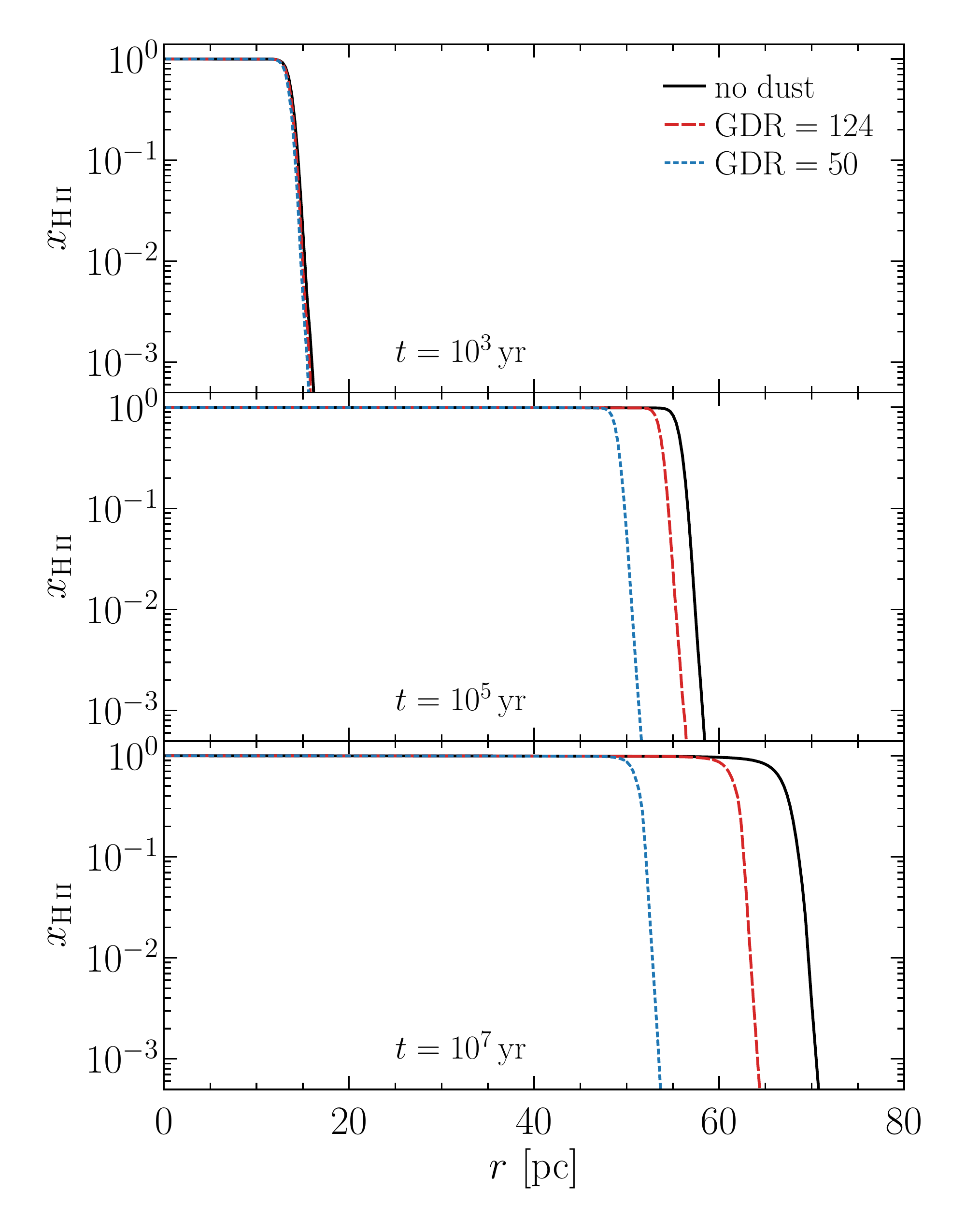}
  \caption{Spherical averages of $x_\HII$ as function of distance from the
    black-body source for the dust free case and GDR values \num{50} and
    \num{124} at simulation times \SI{e3}{yr} (upper panel), \SI{e5}{yr}
    (middle) and \SI{e7}{yr} (lower).}
  \label{fig:bubbles-time}
\end{figure}

\begin{figure}
  \centering
  \includegraphics[width=\linewidth]{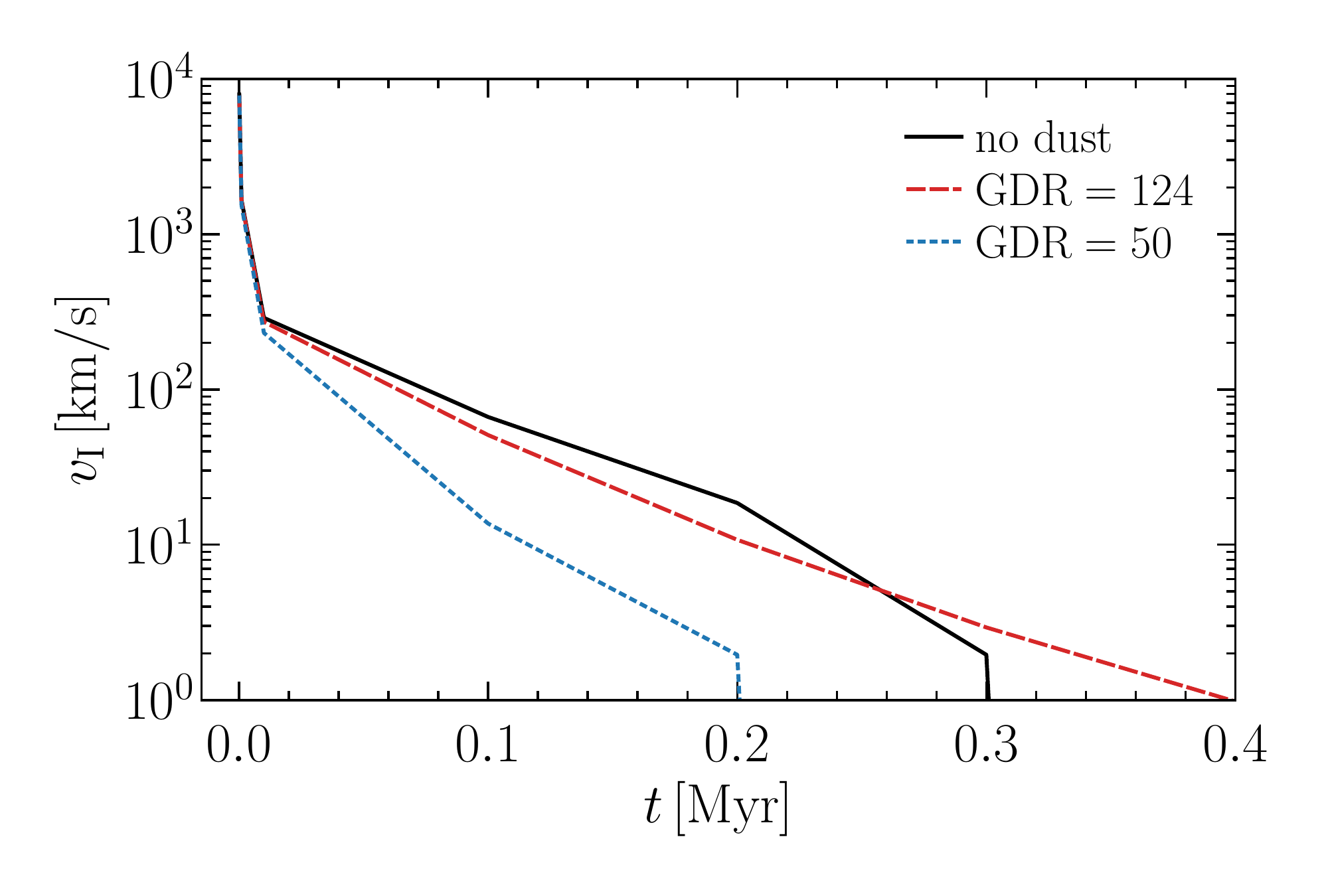}
  \caption{Expansion speed of the \HII\ ionization front as a function of
    simulation time for the dust free case and GDR values \num{50} and
    \num{124}.}
  \label{fig:v_I}
\end{figure}

After discussing the equilibrium configurations found in our simulations, we
now go over to investigating the effect of dust on the time evolution of \HII\
regions. Fig.~\ref{fig:bubbles-time} shows the growth in time of the ionized
bubble at different GDR values, and Fig.~\ref{fig:v_I} shows the expansion
speed of the corresponding ionization fronts as a function of time. While
initially all fronts coincide and propagate at the same speed, slowdown,
dropout and stagnation of fronts in the order of increasing GDR value can be
observed at later times when the photons have to traverse a considerable dust
column in order to reach the front. The slowdown due to dust absorption might
have implications for the overlap of ionized bubbles and consequently for the
percolation of dusty media containing several sources (see also
\S~\ref{sec:rt-dusty-web}). Note that we lack the numerical precision to
accurately compute speeds lower than $\sim\SI{10}{km\per s}$ and that we
attribute the premature speed drop in the dust free case to this fact.

A word of caution concerning all above findings is necessary at this point,
since our limited dust implementation is, for the time being, not directly
coupled to the gas. We work on implementing also grain charging, which results
in grains and gas sharing the free electron population\footnote{Note that in
  general dust will not significantly impact the free electron density in
  ionized regions, since there is far more mass (and therefore electrons to be
  released) in the gas than in the dust. In photon-dominated regions, on the
  other hand, gas-dust reactions can change the electron density
  \citep{Abel2008}.}; this will introduce direct feedback of dust on gas
recombination and cooling, since energetic photo-electrons released from the
grains will provide an additional channel for the radiation field to heat the
gas. Whether grains provide net heating or cooling thus likely depends on the
specific environment being considered \citep[e.g.][]{Weingartner2001b} and is
one the questions we will investigate in future work. Many other aspects of
dust physics can later also be included in the code.

\subsubsection{The impact of PAHs}
\label{sub: PAH-HII}

In this section we quantify the impact of PAH removal on the ideal \HII\
region size of our simple model. Ideally, one should couple the evolution of
the PAH population to the RT in time by directly implementing PAH destruction
and creation processes in the gas chemistry. However, here we take a far
simpler approach and compare simulation runs adopting mixtures with and
without PAHs to assess the maximum possible effect on the spherical \HII\
profiles at equilibrium time.

\begin{figure}
  \centering
  \includegraphics[width=\linewidth]{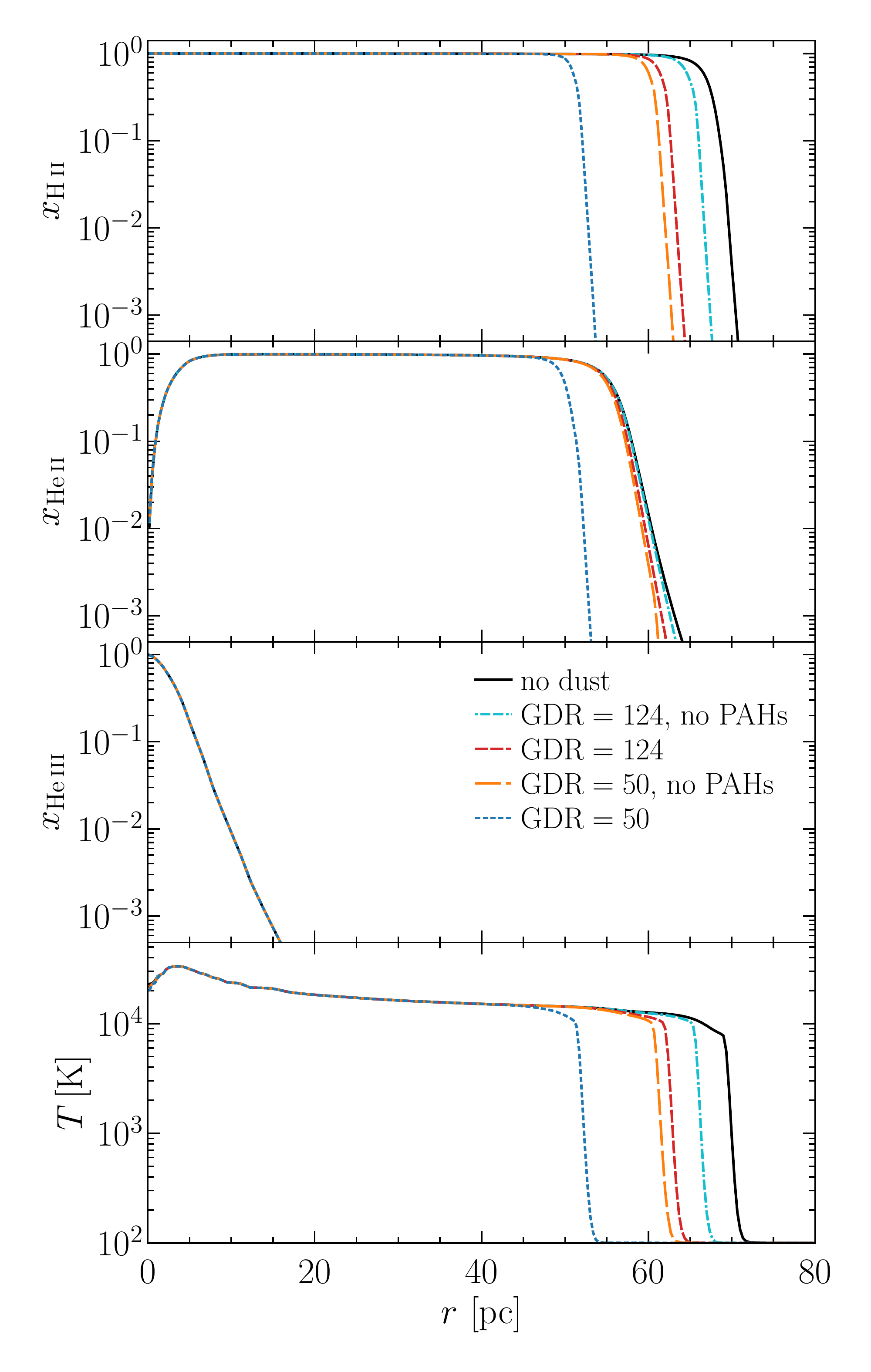}
  \caption{As Fig.~\protect\ref{fig:BBSphAvg}, but showing results including
    (excluding) the PAH contribution to the cross section for
    $\text{GDR}=\num{124}$ in dashed red (dash-dash-dotted cyan) and for
    $\text{GDR}=\num{50}$ in dotted blue (long dashed orange). For reference,
    we also show the dust free case (solid black).}
  \label{fig:noPAH}
\end{figure}

After verifying that the small differences introduced in the cross sections
we computed (see solid black versus dashed red lines in
Fig.~\ref{fig:PAH-dest}) reproduce the profiles obtained using the
Silicate-Graphite-PAH model, we discuss the effects of removing the PAH
component (in the $b_\mathrm{C}=\SI{0.0}{ppm}$ approach) in
Fig.~\ref{fig:noPAH} using the example of our reference case
($\text{GDR}=\num{124}$) and the case of $\text{GDR}=\num{50}$.

In our reference case, the effect of the PAH component can be appreciated
only in the I-fronts of the hydrogen and temperature profiles, which lie
between the full dust and no dust cases as expected, while it is completely
negligible in the ionization fractions of helium. This is in line with the
finding from the previous section that dust absorption leaves the inner
structure of ionized regions unchanged and only affects their fronts. It is
thus interesting, albeit purely coincidental, to note that in the
$\text{GDR}=\num{50}$ case, the removal of the PAH component nullifies the
effects of the dust in the \HeII\ profile, while it provides almost identical
fronts to the full dust $\text{GDR}=\num{124}$ case in both hydrogen and
temperature profiles.

We also tested the PAH impact on the I-front velocity profiles of the \HII\
region, finding that in the $\text{GDR}=\num{124}$ case the absence of the
PAH component results in a front speed equivalent to the dust free case
(black line in Fig. \ref{fig:v_I}). For $\text{GDR}=\num{50}$, instead, the
front expands at a rate similar to that of full dust $\text{GDR}=\num{124}$
case.

We have verified that the above conclusions hold qualitatively also if the
source has a black body spectral distribution peaking at $\sim \SI{17}{eV}$,
where PAHs provide the highest absorption, finding no evidence for a strong
dependence of our results on the spectral shape.

The impact of the PAH component will be explored again in
\S~\ref{sec:rt-dusty-web} to investigate a more complex scenario involving a
realistic cosmic web and a combination of sources.

\subsection{RT through a dusty cosmological volume}
\label{sec:rt-dusty-web}

Here we examine the combined effect of several stellar-type sources on a
small cosmic web artificially polluted with dust. This allows us to
investigate if a diffuse dusty medium can affect the global pattern of
overlapping \HII\ regions with respect to a dust free medium and thus change
the morphology and timing of the resulting small scale reionization driven by
galactic sources.

To mimic the above process, we adopt the simulation set-up of Test~4 of the
\textit{Cosmological Radiative Transfer Comparison Project}
\citep{Iliev2006}, which was later adapted by \cite{Graziani2013} to
investigate the effects of atomic-metal pollution. We only briefly describe
the RT configuration here and refer to the original papers for more
details. The simulation has a box size
$L_\mathrm{b}=0.5h^{-1}\,\mathrm{cMpc}$ ($h=0.72$), a grid resolution
$N_\mathrm{c}=128$ (corresponding to a spatial resolution of
$4h^{-1}\,\mathrm{ckpc}$), a volume-averaged gas density
$\left<n_\mathrm{gas}\right>\approx \SI{2.3e-4}{cm^{-3}}$ and
$t_\mathrm{f}=\SI{0.5}{Myr}$ starting at redshift $z=9$. The cosmological
density evolution is not accounted for. \num{16} hard
($T_\mathrm{bb}=\SI{e5}{K}$) black-body type sources with emissivities
$\dot{N}\sim\SI{e52}{s^{-1}}$ are distributed throughout the volume, and
their spectra are sampled using \num{21} frequencies. \num{e8} photon packets
are emitted per source, resulting in sub-permille variations of volume
averages when compared to test runs with \num{2e8} packets.

\begin{figure}
  \centering
  \includegraphics[width=\linewidth]{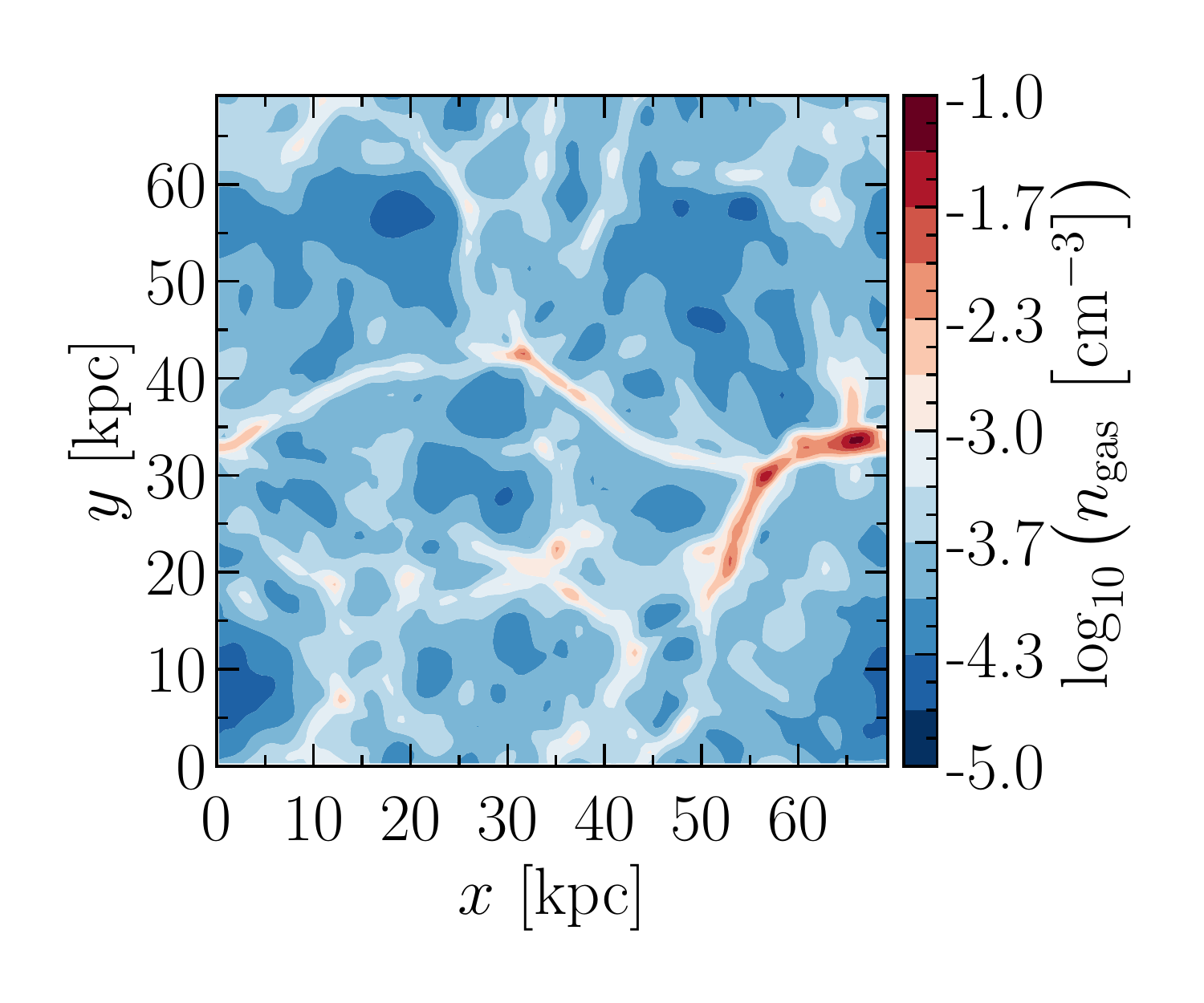}
  \caption{Contour map of the gas number density in a slice cut of the
    simulated cosmological volume.}
  \label{fig:CWebGas}
\end{figure}

The gas number density in a slice cut of the simulation volume is shown in
Fig.~\ref{fig:CWebGas}. The plane of the slice cut has been selected to
intercept one of the most luminous sources as also done in
\cite{Graziani2013}.
\begin{figure*}
  \centering
  \includegraphics[width=0.8\textwidth]{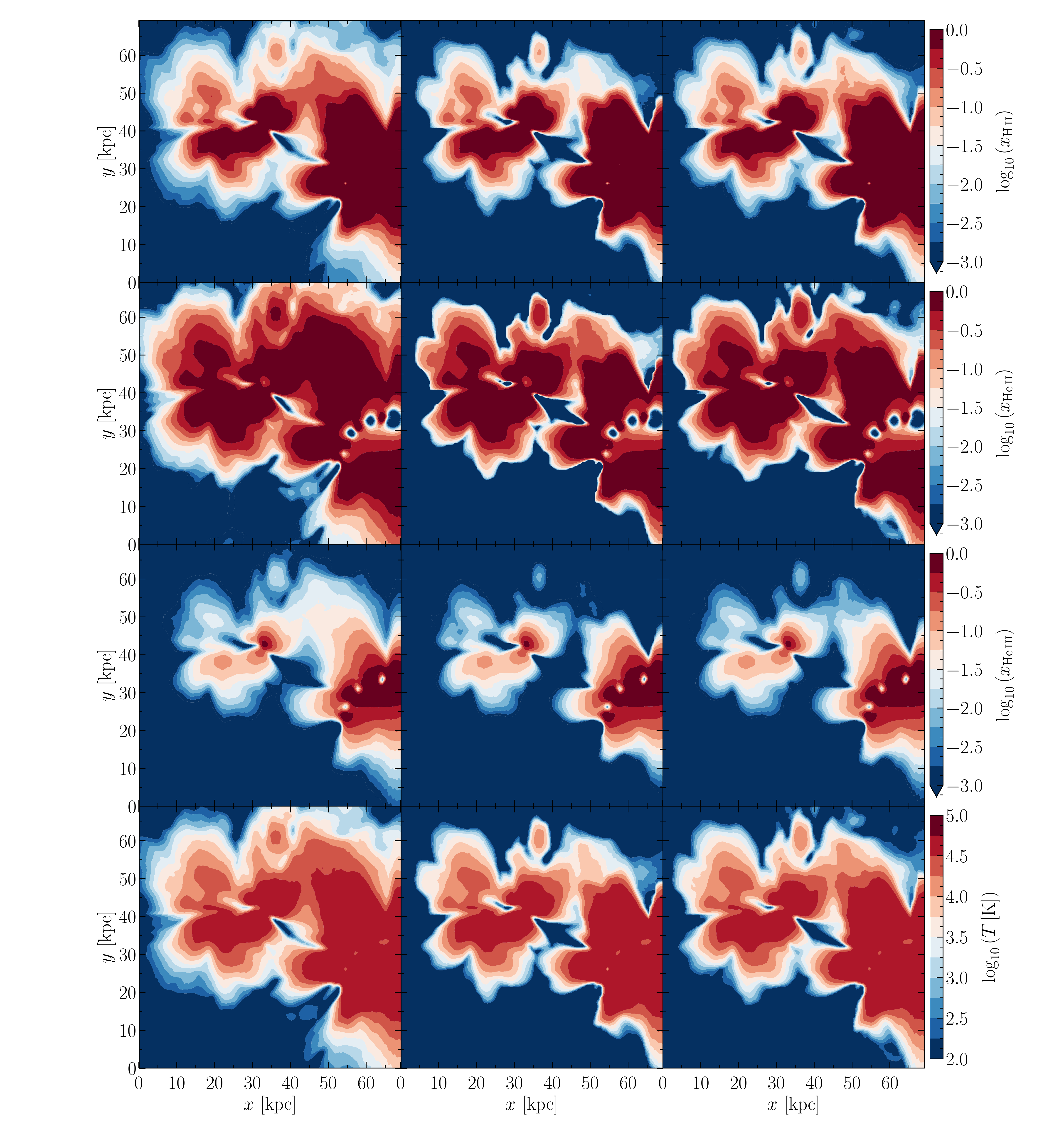}
  \caption{Contour maps of a slice cut through the simulation volume at time
    $t = \SI{5e4}{yr}$. From top to bottom the panels show: \HII\ fraction,
    \HeII\ fraction, \HeIII\ fraction and temperature. The columns show dust
    free results, results with full dust pollution and results with PAH free
    dust pollution from left to right .}
  \label{fig:CWebFinal}
\end{figure*}

We pollute the medium with dust by simply scaling the gas mass in each cell
with a GDR of \num{124}. Note that this, though guided by literature results,
is an essentially arbitrary choice, since the abundance of dust at $z=9$ is
observationally unconstrained. Moreover, the restrictions concerning dust
evolution and its consequences on extinction as discussed in
\S~\ref{sec:silic-graph-pah} apply. Our simulation can therefore only give
first qualitative results assuming early and efficient dust pollution, which
is not inconceivable at this scale. To emulate a decrease of GDR with
increasing distance from galactic centres \citep[e.g.][]{Giannetti2017}, we
also perform a test using overdensity
($\Delta = n_\mathrm{gas}/\left<n_\mathrm{gas}\right>$) dependent GDRs. We
find that the results shown in the following remain largely unaffected when
using $\text{GDR}=\num{124}$ in cells with $\Delta> \num{100}$,
$\text{GDR}=\num{500}$ for $\num{10} < \Delta < \num{100}$ and
$\text{GDR}=\num{1000}$ (\num{e6}) in all other cells, even though doing so
decreases the total dust mass in the simulation volume to
$\sim\SI{15}{\percent}$ ($\sim\SI{5}{\percent}$) with respect to the constant
GDR case. In future studies we plan to investigate initial conditions
generated by hydrodynamic simulations that treat dust in a self-consistent
manner. Dust abundances are, however, also highly uncertain from a
theoretical/numerical point of view, especially on the circumgalactic medium
scale, where they strongly depend on the feedback model adopted in a
simulation.

In Fig.~\ref{fig:CWebFinal} we present contour maps of the plane shown in
Fig.~\ref{fig:CWebGas} at the time $t = \SI{5e4} {yr}$, comparing results
obtained without dust (left), with the full dust mixture (middle) and with PAH
free dust (right). The volume averaged H ionization fractions are \num{0.14},
\num{0.12} and \num{0.13}, respectively. First note that, owed to the shape of
the spectrum of the sources, \HeII\ regions tend to be more extended than \HII\
regions \citep[see also][Fig.~3 and \S~5.2.3]{Graziani2013}. Moreover, it is
difficult to investigate the shape of a single ionized region as there are
multiple sources in the main overdense filament seen in Fig.~\ref{fig:CWebGas},
and at the chosen time several ionized bubbles already overlap. In the top
panels, where ionized hydrogen is shown, we immediately note the sharpening of
ionization fronts. The degree of \HII\ region overlap is consequently reduced
by the presence of dust, changing the spatial distribution of the ionized
regions in the polluted medium. An even more pronounced effect is visible in
the $x_\HeII$ and $x_\HeIII$ patterns, where the ionizing flux subtracted by
the dust also favours a faster helium recombination. In the bottom panel, the
gas temperature distribution is shown. The indirect effects of dust absorption
on the gas temperature are evident in the form of a less progressive transition
between cold ($T \sim \SI{e2}{K}$) and hot ($T \sim \SI{e4}{K}$) gas. We call
particular attention to two points: first, note how at coordinates
(\SI{45}{kpc}, \SI{45}{kpc}) in the second row the highly ionized regions
overlap in the dust free case, whereas they hardly touch in the dust polluted
case. Second, one can appreciate at (\SI{40}{kpc}, \SI{35}{kpc}) how
self-shielding neutral pockets can survive longer in the presence of dust.

The effects of removing the PAH component correspond to what one would expect
from our study of single \HII\ regions, i.e. without PAHs the I-fronts fall
between those of the dust free and full dust cases. We do therefore not
discuss the right column in more detail.

\begin{figure*}
  \centering
  \includegraphics[width=\textwidth]{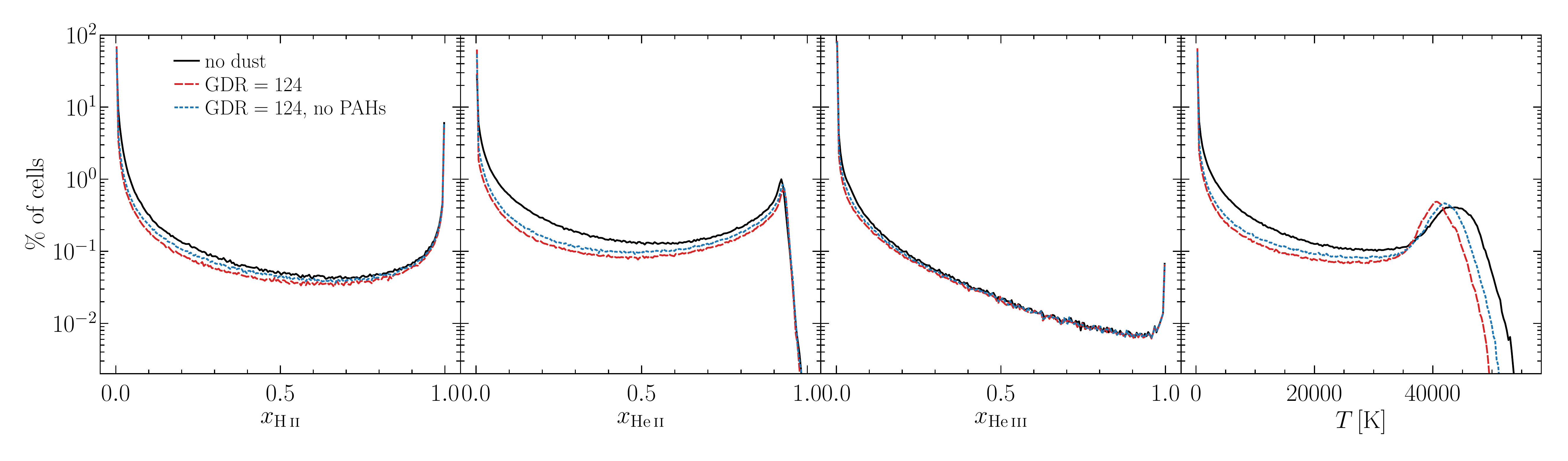}
  \caption{Distributions of all cells in the simulation volume in (left to
    right) $x_\HII$, $x_\HeII$, $x_\HeIII$ and $T$ at $t = \SI{5e4}{yr}$ for
    the dust free (solid black line), dust polluted (dashed red) and PAH free
    (dotted blue) runs. The cells were distributed onto \num{200} bins.}
  \label{fig:CWebHist}
\end{figure*}

Fig.~\ref{fig:CWebHist} gives a quantitative view of the differences between
our dust free, dust polluted and PAH free results. It shows the distribution
of all cells in the simulation volume at $t = \SI{5e4}{yr}$ for $x_\HII$,
$x_\HeII$, $x_\HeIII$ and $T$. As one would expect from the above discussion,
dust increases the number of neutral cells, reduces the amount of weakly
ionized cells and leaves the number of highly ionized cells nearly
unchanged. Moreover, it shifts the peak corresponding to hot, ionized cells
in the temperature distribution towards lower temperatures and reduces the
amount of cells heated to $\sim$\SIrange{e3}{3e4}{K}.

\begin{figure}
  \includegraphics[width=\linewidth]{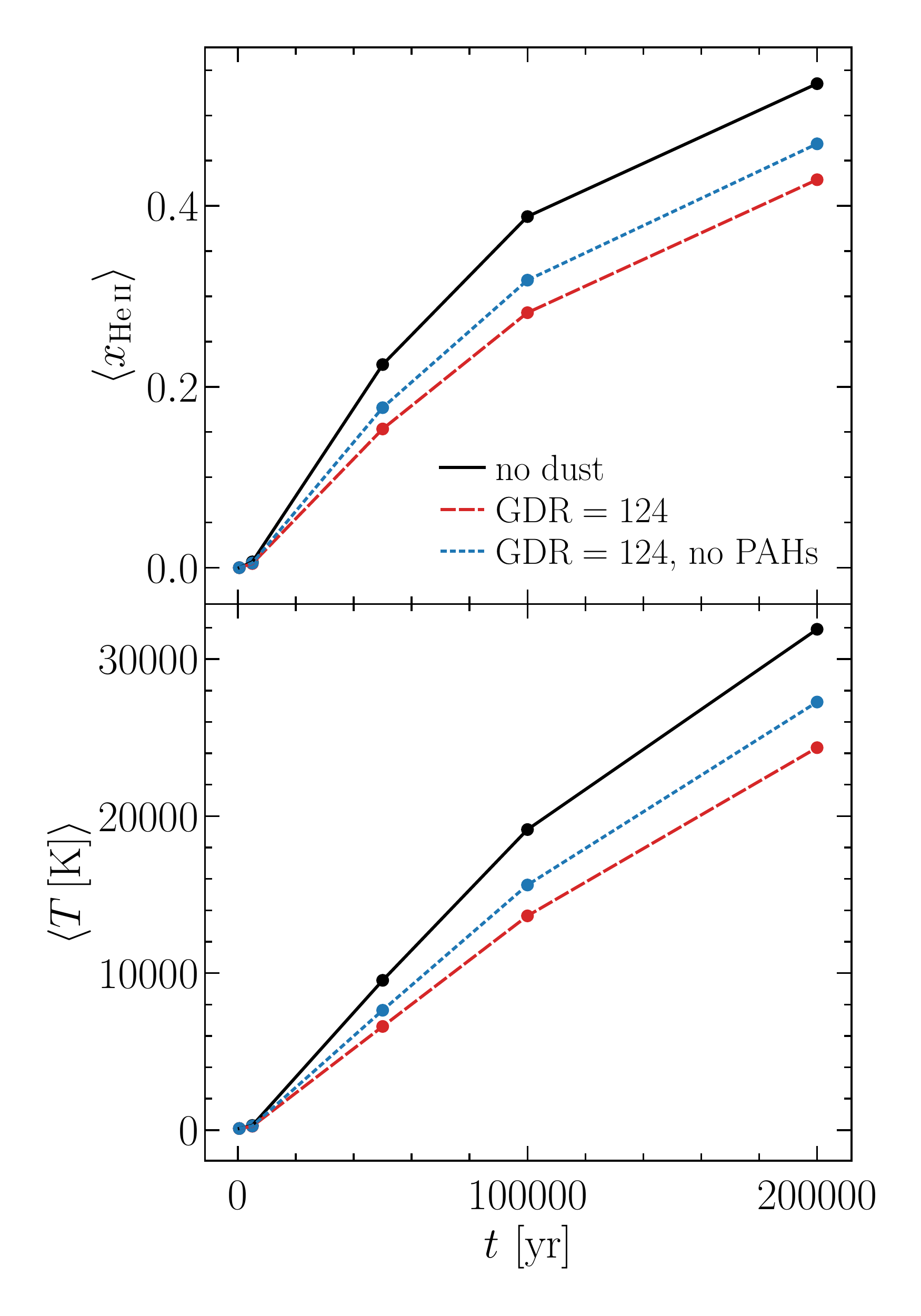}
  \caption{Time evolution of the volume averaged $x_\HeII$ (top panel) and
    $T$ (bottom) for our dust free (solid black line), $\text{GDR}=\num{124}$
    (dashed red) and PAH free (dotted blue) run.}
  \label{fig:time-evol}
\end{figure}

Owed to the location of the peak of dust absorption, $x_\HeII$ and $T$ are
most affected in volume averages. In Fig.~\ref{fig:time-evol} we therefore
show the corresponding time evolution. At \SI{0.2}{Myr}, the dust free case
has a volume averaged $x_\HeII$ ($T$) of \num{0.54} (\SI{3.2e4}{K}), which is
reduced by \SI{20}{\%} (\SI{25}{\%}) in the full dust case and \SI{12}{\%}
(\SI{16}{\%}) in the PAH free case.

To investigate the dependence on the spectral shape of all the above
findings, we performed simulation runs in which we used
$T_\mathrm{bb}=\SI{4e4}{K}$ black-body spectra for the sources, while keeping
all the other parameters fixed. The associated shift in the spectrum peak
primarily results in less helium ionization and in a lower average gas
temperature, but qualitatively the above observations still hold.

Dust can offer significant absorption at ionizing energies. The simple cases
studied here show sensitive alterations of size, shape and time evolution of
ionized regions: ionization fronts get sharpened, the size of fully ionized
regions is reduced changing their degree of overlap and the gas temperature is
lowered. These effects can certainly have a strong impact on small scale RT
simulations, where dust is known to be present and its spatial extent can be
resolved. It should be considered, however, that other physical processes
currently not included in our implementation can be important. In particular,
dust photo-heating is likely to have a strong impact on the gas temperature and
will have to be accounted for in the future. Further observational and
numerical investigations of the spatial distribution, abundance and chemical
properties of dust are required in order to critically examine the assumptions
made, to determine the appropriate physical set-up of future RT simulations and
to precisely quantify the impact of absorbing grains on the final ionization
and temperature patterns created in the many different environments of both the
ISM and the IGM.

\section{Conclusions}
\label{sec:conclusions}

In this work we have extended the Monte Carlo Radiative Transfer (RT) code
for ionizing ($h\nu\geq \SI{13.6}{eV}$) radiation \crash\ by a dust module
that, at the moment, accounts for the absorption of radiation by dust. We
have used this new implementation to study the formation of idealized \HII\
regions by point-like stellar sources and found that including dust leaves
their inner ionization and temperature structure largely unchanged. As
expected \citep[e.g.][]{Draine2011}, however, it can result in a significant
reduction in size. For monochromatic sources, our code predicts \HII\ region
sizes in accordance with semi-analytic solutions (see
Appendix~\ref{sec:comparison-petrosian}). Furthermore, we have performed a
first step towards investigating the effect of dust on small scale
reionization by performing RT on a small cosmological volume artificially
polluted with dust and containing several black-body sources. Here we find
that dust primarily sharpens ionization fronts and slows down the overlap of
intergalactic ionized bubbles. This, naturally, also results in a decrease in
volume of hot $(\sim\SI{e4}{K})$ gas. There seems to be only a weak
dependence of these results on the gas-to-dust mass ratio used in the
underdense regions of the cosmological volume.

The next project we plan to approach using the implementation presented here
is investigating non-equilibrium ionization by Active Galactic Nuclei (AGN),
complementing the work done in \cite{Kakiichi2017} with dust. Such sources
allow studying X-ray effects since they emit large numbers of photons at high
energies, where the dust cross section gains importance in comparison with
the gas cross section. In the future we plan to simulate RT in
post-processing on gas and dust distributions self-consistently computed by
hydrodynamics codes. This is a crucial step to improve on the simple dust
distributions used here. Moreover, the dust physics in our model will be
further developed to include effects such as grain charging, heating and
destruction.

\section*{Acknowledgements}
The authors are very grateful to Bruce Draine and Joseph Weingartner for
their valuable and rapid answers to questions regarding their work on the
Silicate-Graphite-PAH dust model. We also thank Simone Bianchi and Andrea
Ferrara for their insightful comments on the manuscript.

Finally, we would like express our gratitude to the anonymous referee for
reading the manuscript with great care and for providing challenging comments
that resulted in a constructive discussion and substantial improvements.

%%%%%%%%%%%%%%%%%%%%%%%%%%%%%%%%%%%%%%%%%%%%%%%%%%


\bibliographystyle{mnras}
\bibliography{main.bib}

%%%%%%%%%%%%%%%%%%%%%%%%%%%%%%%%%%%%%%%%%%%%%%%%%%

%%%%%%%%%%%%%%%%% APPENDICES %%%%%%%%%%%%%%%%%%%%%

\appendix

\section{Comparison with analytic solution}
\label{sec:comparison-petrosian}
\cite{Stroemgren1939}, assuming strongly idealized conditions, developed a
well known approximate analytic expression for the hydrogen ionization
fraction as a function of distance from a star in ionization equilibrium with
the gas surrounding it \citep[e.g.][]{Draine2011}, which
\cite{Petrosian1972}\footnote{Note that there is a typo in Eq. (7) of the
  paper. $1-y^{3/4}$ should read $1-y^3/4$.} later extended to include dust
in the form of a homogeneously distributed absorber of
radiation. \cite{Raga2015} improved on the \cite{Petrosian1972} solution with
a more complex approach. In the regime we investigate below, both agree very
well with the exact (up to numerical integration errors) solution, obtained
by integration of the differential equation describing the change in ionizing
flux with distance from the central star \citep[e.g.][]{Raga2015}. As the gas
density decreases, the agreement tends to worsen, with \cite{Petrosian1972}
predicting sharper ionization fronts \citep[see][Fig.~3]{Raga2015}.

\begin{figure}
  \centering
  \includegraphics[width=\linewidth]{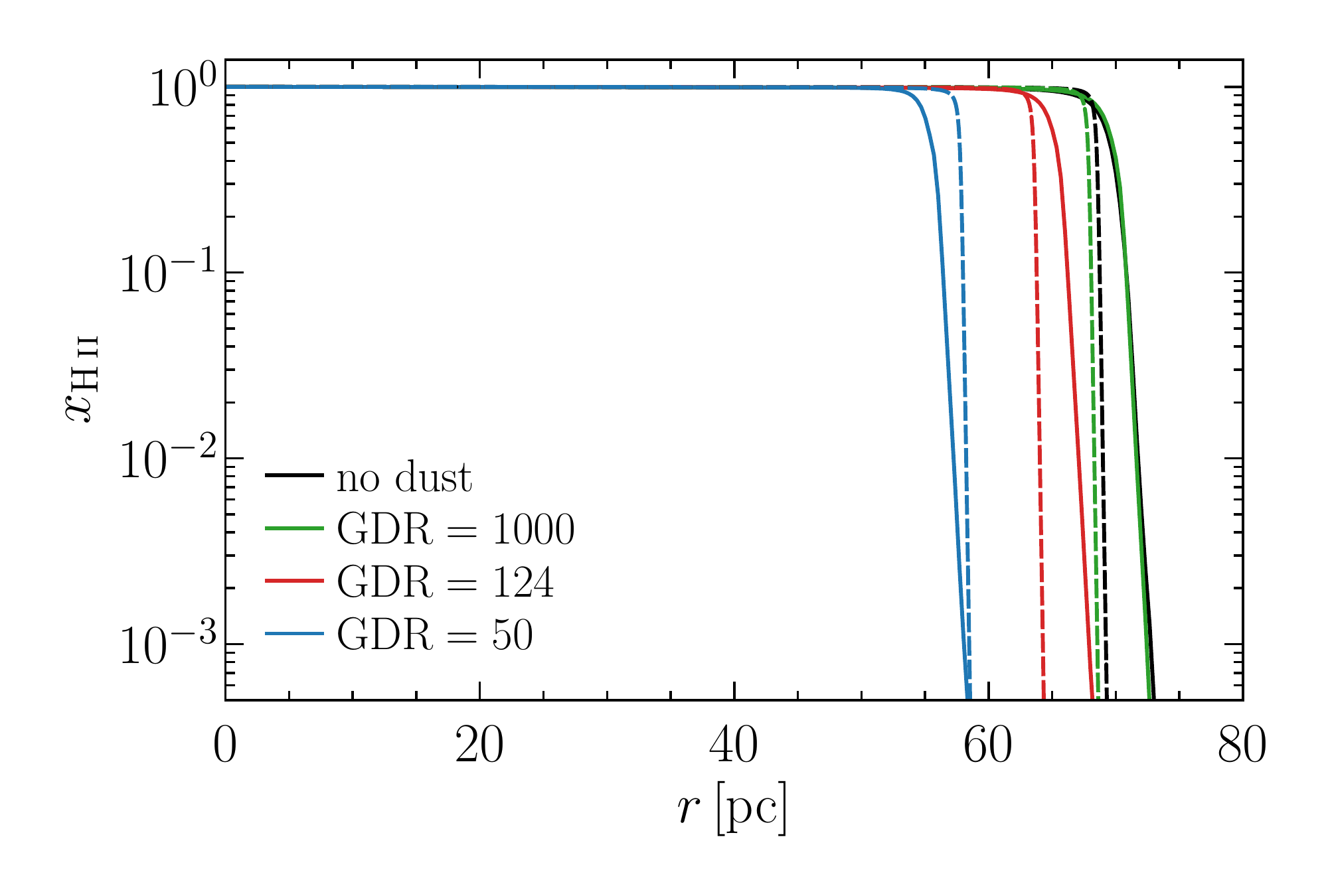}
  \caption{Hydrogen ionization fraction in equilibrium as a function of
    distance from a monochromatic source for GDR values \num{1000}, \num{124}
    and \num{50}. The solid lines are \crash\ results (see text for details),
    and the dashed lines show corresponding numerical integration results
    computed following \protect\cite{Raga2015}.}
  \label{fig:petrosian}
\end{figure}
In Fig.~\ref{fig:petrosian} we show $x_\HII$ as computed by \crash\ (solid
lines) as well as the corresponding solutions of the numerical integration
(dashed lines) as discussed in \cite{Raga2015}. The simulations shown in this
figure are set up as described in \S~\ref{sec:stroemgren-sphere}, with
$n_\mathrm{gas}=\SI{1}{cm^{-3}}$ and various GDR values. For a meaningful
comparison, we have made the following changes:
\begin{itemize}
\item He abundance is set to zero;
\item the source emits monchromatic photons at \SI{13.6}{eV};
\item the gas temperature is set to \SI{e4}{K} in the entire simulation
  volume and kept constant during the simulation;
\item case~B recombination is enabled.
\end{itemize}
As our scheme is time dependent, we must compare the ionization structure
obtained once an equilibrium configuration is reached to a solution obtained
explicitly assuming equilibrium. Despite this, the agreement of the results
is remarkably good in a wide range of gas number densities (due to the
limited space available, we do not show the tests we performed at number
densities other than \SI{1}{cm^{-3}} here) and values of GDR. In the cases
without dust, $\text{GDR}=\num{1000}$ and \num{124}, the location of the
I-front is very well reproduced by \crash , while intrinsic differences
between numerical algorithm and analytic solution can induce a wider/narrower
declining front. For $\text{GDR}=\num{50}$, \crash\ slightly overpredicts
absorption, but remains in good agreement. It should be noted that an exact
comparison between (semi-)analytic solutions and predictions of numerical RT
codes is always difficult beyond the I-front position for a number of
reasons. First, each numerical method has its own intrinsic issues in
sampling the radiation field: the long-characteristic method implemented in
\crash , for example, requires a large number of photon packets to correctly
cover the solid angle of emission of isotropic sources. Without the
appropriate angular coverage adopted here and in \S~\ref{sec:rt-dusty-web},
at large distances from the source the declining fronts would show a spiky
behaviour due to the geometrical separation of the few rays escaping the
I-front. This computational issue is solved in short-characteristic
approaches but at the price of having less radial coherence in the
propagating packets. Second, and more important, analytic solutions often
rely on over-simplifications that advanced numerical codes have been build to
encompass (e.g. single frequency photons, fixed temperature, missing
collisional/secondary ionization, etc.). The collisional ionization term, for
example, is generally neglected in the analytic estimates while fully
implemented in \crash\ as function of the evolving temperature and cannot be
switched off during the simulation. Depending on the assumed gas density and
temperature this term could induce significant differences in the way the
ionization fronts decline at imposed, constant temperature. Finally, note
that apart from these idealized set-ups, all the simulations shown in this
paper adopt consistent temperature and ionization computations as well as
appropriate numbers of photons packets to satisfy Monte Carlo convergence of
the results and to minimize the intrinsic numerical noise mentioned above.

From further tests we conducted following the above findings, we conclude
that for $n_\mathrm{gas} \leq \SI{1}{cm^{-3}}$ and $\text{GDR}\geq \num{50}$,
and for $n_\mathrm{gas} \leq \SI{0.1}{cm^{-3}}$ and $\text{GDR}\geq \num{10}$
our scheme shows good agreement with the numerical integration, while caution
should be exercised for higher gas and dust densities. In those regimes the
time independent dust optical depth significantly alters the time evolution
of the I-front and reduces its final equilibrium radius by up to a factor of
a few with respect to the numerical integration. We are further investigating
this and believe careful comparisons with other time dependent RT schemes
accounting for dust would provide helpful insights.

\section{Modelling the effect of PAH destruction on the dust cross section}
\label{sec:nopah-dust-model}

Since no full model of the processing of PAHs in \HII\ regions is available,
it is not clear how one should modify a given grain size distribution and
thus the resulting cross section (see \S~\ref{sec:dust-PAH}) of a dust
mixture to account for it. We explore three different approaches here,
presented in detail in the following. Note that we never change the
distribution of silicate grains.

As described in \S~\ref{sec:silic-graph-pah}, the Silicate-Graphite-PAH model
does not feature a separate PAH population, but PAHs are rather modelled as a
part of the carbonaceous grain population. Specifically, LD01 introduce in
Eq.~(3) a PAH ``weight'' $\xi_\mathrm{PAH}(a)$, where $a$ is the effective
grain radius, which they use to make a transition from graphite to PAHs in the
absorption cross section of carbon grains (see Eq.~(2) of LD01). The total
optical depth per distance $l$ due to carbon dust with cross section
$\sigma^\mathrm{carb}(a, \nu)$ and grain size distribution
$\diff n^\mathrm{carb}/\diff a$ is then computed as:
\begin{equation}
  \label{eq:tau-d}
  \begin{split}
    \frac{\tau^\mathrm{carb}(\nu)}{l}
    &= \int\!\diff a \tder{n^\mathrm{carb}}{a}\sigma^\mathrm{carb}(a, \nu) \\
    &= \int\!\diff a
    \tder{n^\mathrm{carb}}{a}(\xi_\mathrm{PAH}\sigma^\mathrm{PAH}(a, \nu)) +
    (1-\xi_\mathrm{PAH})\sigma^\mathrm{gra}(a, \nu))\,,
  \end{split}
\end{equation}
with PAH ($\sigma^\mathrm{PAH}(a, \nu)$) and graphite
($\sigma^\mathrm{gra}(a, \nu)$) cross sections as defined in WD01.

In our first approach, which we refer to as ``split'', we propose to split
the carbon grain population into a PAH and a graphite population, i.e. we
define grain size distributions
\begin{equation}
  \label{eq:n-pah}
  \tder{n^\mathrm{PAH}}{a}=\xi_\mathrm{PAH}(a)\tder{n^\mathrm{carb}}{a}
\end{equation}
and
\begin{equation}
  \label{eq:n-gra}
  \tder{n^\mathrm{gra}}{a}=(1-\xi_\mathrm{PAH}(a))\tder{n^\mathrm{carb}}{a}\,.
\end{equation}
With these definitions, the total carbon dust optical depth becomes:
\begin{equation}
  \begin{split}
    \frac{\tau^\mathrm{carb}(\nu)}{l}
    &= \int\!\diff a\left(
      \tder{n^\mathrm{PAH}}{a}\sigma^\mathrm{PAH}(a, \nu) +
      \tder{n^\mathrm{gra}}{a}\sigma^\mathrm{gra}(a, \nu))\right) \\
    &=\frac{\tau^\mathrm{PAH}(\nu)}{l} + \frac{\tau^\mathrm{gra}(\nu)}{l}\,.
  \end{split}
\end{equation}
To account for \HII\ region processing, we then set
$\diff n^\mathrm{PAH}/\diff a$ to zero when computing cross sections.

\begin{figure}
  \centering
  \includegraphics[width=\linewidth]{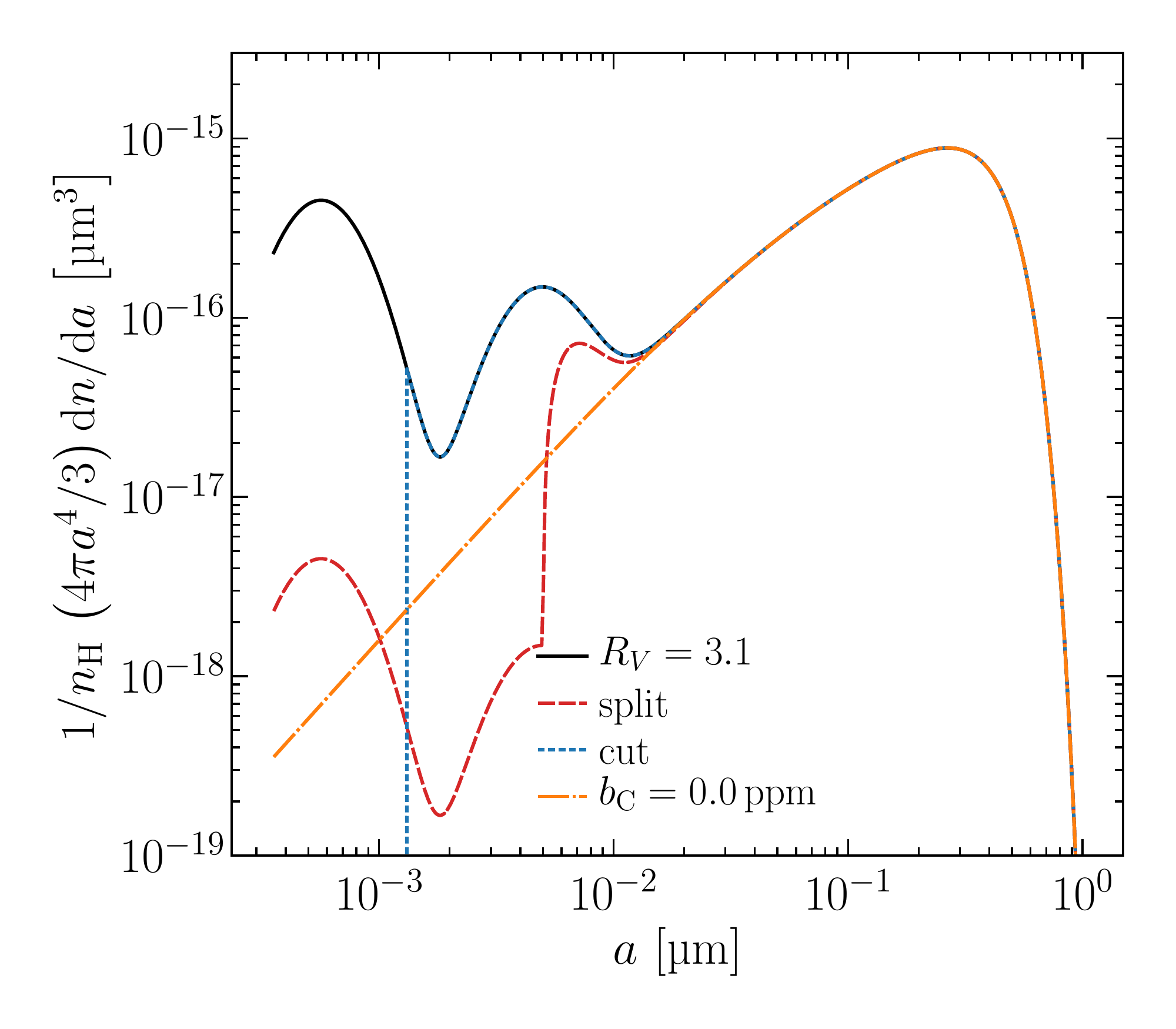}
  \caption{Volume distributions for different carbonaceous grain
    populations. The black solid line shows the unmodified distribution for
    $R_V=\num{3.1}$ and $b_\mathrm{C}=\SI{60}{ppm}$ as in WD01, with the
    other lines showing our approaches to modelling \HII\ region processing
    of dust as discussed in the text. The split, cut and
    $b_\mathrm{C}=\SI{0.0}{ppm}$ approach result in a dust (including
    silicate grains) mass density reduction by \SI{6.1}{\%}, \SI{4.7}{\%} and
    \SI{6.3}{\%} relative to the unmodified distribution,
    respectively. Cf. \protect\cite{Draine2007}.}
  \label{fig:distros}
\end{figure}

The motivation for this approach lies in excluding the contribution to the
cross section attributed to PAHs in the Silicate-Graphite-PAH model. This
does, however, result in a size distribution (see dashed red line in
Fig.~\ref{fig:distros}) unlikely to represent physical reality, since it
features a kink at \SI{50}{\angstrom} and rises again towards the lower size
limit, while the smallest PAHs are expected to be the most susceptible to
photo-dissociation \citep[e.g.][]{Bocchio2012}.

\cite{Draine2007} use $q_\mathrm{PAH}$, the mass fraction of dust contained
in carbonaceous particles with fewer than \num{1000}~C atoms, to quantify the
PAH content of a given grain size distribution. Although the PAH destruction
efficiency is unlikely to feature a sharp transition specifically at
\num{1000}~C atoms (roughly \SI{13}{\angstrom} in effective radius), we base
our second approach, referred to as ``cut'' (dotted blue line in
Fig.~\ref{fig:distros}), on this definition and sharply cut the carbon grain
size distribution here when computing cross sections for \HII\ region
processed dust. Otherwise the carbon grains are treated as in the original
model.

In the third approach we set $b_\mathrm{C}=\SI{0.0}{ppm}$ while keeping all
other parameters of the size distributions fixed. This removes the two
log-normal distributions that WD01 add to the power-law distribution for the
carbonaceous grain population, significantly reducing the number of small
carbon grains. We then proceed as in the cut case described above. This
approach is motivated by the fact that the MW3.1\_00 model from WD01
\citep[see also][Table 3]{Draine2007}, which also does not include log-normal
distributions, is used to fit dust emission from regions of expected low PAH
content (\citealt{Draine2007}; Chastenet et al., in prep.). It produces the
most plausible distribution (dash-dotted orange line) and predicts
essentially the same cross section as our split approach (see
Fig.~\ref{fig:PAH-dest}).

To explore the effect of our modified grain size distributions, knowledge of
the absorption cross section as a function of photon energy and grain size is
required. We can therefore not rely on the published tables of
population-averaged cross sections as in the main text of the paper. Instead,
we follow the prescriptions given in WD01, LD01 and \cite{Draine2003a,
  Draine2003b, Draine2003c} to directly compute cross sections. Although we
plan to investigate the transfer of X-rays in the future, for the purposes of
this paper we can restrict our attention to the energy range
\SIrange{13.6}{160}{eV}, since none of the spectra we used here extends
beyond this range. As stated in \S~3 of \cite{Draine2003c}, the Wiscombe Mie
theory code was used to compute graphite and silicate cross sections for
$x=2\pi a/\lambda < \num{2e4}$, with $\lambda$ the photon wavelength. For
$h\nu=\SI{160}{eV}$ and $a=\SI{10}{\micro m}$ (upper limit of the grain size
distribution), $x\approx \num{8100}$, so that we use Mie
theory\interfootnotelinepenalty=10000\footnote{We obtained a version of the
  Wiscombe code at \url{scatterlib.wikidot.com/mie} and downloaded
  \cite{Draine2003b} dielectric functions from
  \url{www.astro.princeton.edu/~draine/dust/dust.diel.html}.} in the entire
energy range. For the PAH cross section we use Eqs.~(5) to (11) from LD01,
although only Eqs.~(5), (6) and (7) are relevant for our energy range. Note
that in this regime the cross sections of neutral and ionized PAHs are
identical. Cross sections for energies up to $\sim\SI{31}{eV}$ as a function
of grain size are also provided by
\href{http://www.ias.u-psud.fr/DUSTEM/}{DustEM}~\citep{Compiegne2010} in
precomputed tables, which can alternatively be used as input for our
framework. We verified that using these data gives results in good agreement
with those from the direct Mie computation. The framework used for these
computations will be made publicly available on
\href{https://github.com/findesgh/cosmic_dustbox}{GitHub}.

%%%%%%%%%%%%%%%%%%%%%%%%%%%%%%%%%%%%%%%%%%%%%%%%%%

% Don't change these lines
\bsp	% typesetting comment
\label{lastpage}
\end{document}